\documentclass[fleqn,usenatbib]{mnras}

\usepackage{newtxtext,newtxmath}
\usepackage[T1]{fontenc}

\DeclareRobustCommand{\VAN}[3]{#2}
\let\VANthebibliography\thebibliography
\def\thebibliography{\DeclareRobustCommand{\VAN}[3]{##3}\VANthebibliography}

\newcommand{\ve}[1]{\boldsymbol{#1}}

\usepackage{graphicx}	

\usepackage{amsmath}	
\usepackage{amssymb}	
\usepackage{dsfont}
\usepackage{algorithm}
\usepackage{algpseudocode}
\usepackage[mathscr]{euscript}

\title[Optimizing Non-Differentiable Simulators]{Learning the Universe: Learning to Optimize Cosmic Initial Conditions with Non-Differentiable Structure Formation Models}

\author[L. Doeser, M. Ata, J. Jasche]{
Ludvig Doeser$^{1}$\thanks{E-mail: ludvig.doeser@fysik.su.se},
Metin Ata$^{1,2}$,
Jens Jasche$^{1}$
\\
$^{1}$The Oskar Klein Centre, Department of Physics, Stockholm University, AlbaNova University Centre, SE 106 91 Stockholm, Sweden \\
$^{2}$Center for Gravitational Physics and Quantum Information, Yukawa
Institute for Theoretical Physics, Kyoto University, Kyoto 606-8502, Japan
}

\date{Accepted XXX. Received YYY; in original form ZZZ}
\pubyear{2025}

\begin{document}
\label{firstpage}
\pagerange{\pageref{firstpage}--\pageref{lastpage}}
\maketitle

\begin{abstract}
Making the most of next-generation galaxy clustering surveys requires overcoming challenges in complex, non-linear modelling to access the significant amount of information at smaller cosmological scales. Field-level inference has provided a unique opportunity beyond summary statistics to use all of the information of the galaxy distribution. However, addressing current challenges often necessitates numerical modelling that incorporates non-differentiable components, hindering the use of efficient gradient-based inference methods. In this paper, we introduce \textit{Learning the Universe by Learning to Optimize} (\texttt{LULO}), a gradient-free framework for reconstructing the 3D cosmic initial conditions. Our approach advances deep learning to train an optimization algorithm capable of fitting state-of-the-art non-differentiable simulators to data at the field level. Importantly, the neural optimizer solely acts as a search engine in an iterative scheme, always maintaining full physics simulations in the loop, ensuring scalability and reliability. We demonstrate the method by accurately reconstructing initial conditions from $M_{200\mathrm{c}}$ halos identified in a dark matter-only $N$-body simulation with a spherical overdensity algorithm. The derived dark matter and halo overdensity fields exhibit $\geq80\%$ cross-correlation with the ground truth into the non-linear regime $k \sim 1h$ Mpc$^{-1}$. Additional cosmological tests reveal accurate recovery of the power spectra, bispectra, halo mass function, and velocities. With this work, we demonstrate a promising path forward to non-linear field-level inference surpassing the requirement of a differentiable physics model.
\end{abstract}



\section{Introduction}
Harnessing the full potential of data from imminent Stage-IV galaxy surveys, including DESI \citep{DESICollaboration2016}, Euclid \citep{Laureijs2011,Amendola2018}, LSST \citep{LSSTScienceCollaboration2009,LSSTDarkEnergyScienceCollaboration2012,Ivezic2019}, SPHEREx \citep{Dore2014,Dore2018}, and Subaru Prime Focus Spectrograph \citep{Takada2012}, requires increasingly complex data models capable of accurately capturing all features in the observed galaxy distribution. Because galaxies are non-linear gravitationally collapsed objects, a non-linear model is required to establish a causal connection between observational data and cosmological theory. Over the past decades, the cosmology community has therefore been on a program of developing tools to probe increasingly smaller, non-linear scales, where a significantly larger number of observable modes promises an unprecedented amount of constraining information. 

Modern analysis of the large-scale structure of the Universe has been driven by advances in numerical resources, with $N$-body simulations emerging as the most advanced technique for modelling the full non-linear structure formation of the Universe \citep[see reviews by][]{Vogelsberger2020,Angulo2021}. However, incorporating these directly in (iterative) statistical inference frameworks for extracting physics information has not been feasible because of their high computational cost. Efforts were therefore put into developing approximate simulators, which aim to accurately reproduce non-linear structures for a lower cost \citep[see e.g][]{1995A&A...296..575B,Tassev2013,Leclercq2015,Feng2016,Rampf2024}. The development of modern advanced inference technologies was further made possible by algorithmic advancements, most notably the creation of fully differentiable non-linear physics simulators, including differentiable higher-order Lagrangian perturbation theory \citep{Jasche2012}, particle-mesh approaches \citep{Dai2018,Jasche2019,Modi2021,Ll2022} and neural network enhanced models \citep{He2019,AlvesDeOliveira,Bernardini2020,Kaushal,Jamieson2022b, Jamieson2024, Doeser2024, Bartlett2024}.

Performing cosmological inference from data to learn about fundamental physics involves exploration through high-dimensional spaces. These spaces consist of cosmological parameters, determining the average evolution of our Universe, and the field of primordial density fluctuations, from which observations originate. From the requirement of modelling non-linear physics comes the challenge of exploring the corresponding parameter spaces, an often numerically expensive task. To tackle this problem, two approaches are possible:
\begin{enumerate}
    \item Develop fast and differentiable models that accurately approximate a high-fidelity physics model.
    \item Develop optimal search algorithms to maintain full physics predictions in the loop, without requiring differentiability.
\end{enumerate}
In this work, we aim to push forward strategy (ii), while recognizing and leveraging the significant breakthroughs achieved through strategy (i). In particular, differentiable physics simulations allow for rapid exploration of parameter spaces through gradient-based optimization or Hamiltonian Monte Carlo \citep[HMC,][]{Duane1987} sampling facilitated by back-propagation. This has enabled recent data analysis approaches to move beyond perturbative modelling and traditional summary statistics to utilize numerical structure formation models during inference.

Instead of directly modelling the observed galaxy distribution, the task can be reframed as a statistical problem of inferring the initial conditions of the Universe \citep{Jasche2012,Kitaura2013,Wang_2013,Wang2014}. By incorporating physical models of non-linear gravitational structure formation into a Bayesian framework, this approach establishes a direct link between the observed non-linear cosmic structures and the initial density fields from which they evolved \citep[see data applications in e.g.][]{Jasche2015, Lavaux2016, Jasche2019, Kitaura2019, Lavaux2019, Ata2020}. Advancing the forward modelling within these Bayesian frameworks \citep[as in e.g.][]{Ata2014, Schmidt2018, Jasche2019, Bos_2019, Stopyra2023, Doeser2024} is essential for effectively handling the complex, high-dimensional data from upcoming cosmological surveys.

In addition to Bayesian forward modelling, several other approaches to infer the initial conditions have been explored, ranging from maximum-a-posteriori (MAP) estimation \citep{Seljak2017,Feng2018,Horowitz2019,Horowitz2023,Bayer2023} to recent machine learning methods. These include recurrent neural networks \citep{Modi2021a}, auto-regressive models \citep{List2023}, score-based diffusion sampling \citep{Legin2023}, final-to-initial mappings \citep{Jindal2023}, neural-enhanced forward modelling \citep{Modi2018,Doeser2024}, and variational self-boosted sampling \citep{Modi_2023}. Convolutional neural networks have further been widely used to improve small-scale reconstructions \citep[e.g.][]{Shallue2023, Chen2023, Parker2025, Bottema2025, Savchenko2024, Savchenko2025}.

The interest in initial conditions stems from the crucial role that they, along with the laws of physics, play in determining the dynamical evolution and formation of the large-scale structure of galaxies in the Universe. Accurately inferring the initial conditions is thus essential for interpreting observational data, testing cosmological theories, and probing the nature of dark matter and dark energy. Joint inference of initial conditions and cosmological parameters have shown promise, such as for $(\Omega_\mathrm{m},\omega_0)$ \citep{Ramanah2018}, $\sigma_8$ \citep{Kostic2022,Nguyen2024}, $(\Omega_\mathrm{m},\sigma_8)$ \citep{Porqueres2023,Krause2024}, and primordial non-Gaussianties \citep{Chen2024,Andrews2022,Andrews2024,Floss2024}.

A key challenge in accurately modelling observations lies in capturing the connection between matter and haloes or galaxies, commonly referred to as galaxy bias \citep[e.g.][]{Desjacques2016}. For example, \citet{Bartlett2024a} showed that non-linear but local galaxy models are insufficient. As high-dimensional field-level inference methods incorporating non-linear structure formation and perturbative or phenomenological galaxy bias models have now become computationally feasible, the next pressing challenge involves how to integrate non-differentiable models in the data analysis. These include halo population models, galaxy formation models, and hydrodynamical simulations, which offer high-fidelity modelling of, e.g. baryonic physics, massive neutrinos, and feedback-reactions \citep[see, e.g.][]{Villaescusa-Navarro2020a,Schaye2023,Pakmor2023}. Currently, no differentiable algorithms exist for these models, and the implementation and testing of these methods for science-ready applications will require significant development effort and resources to be invested by the community. This is also true for efforts to circumvent traditional non-differentiable models, such as fast differentiable bias models \citep{Modi2018, Kodi_Ramanah_2019, Charnock2020, Ding2024, Pandey2024a, Pandey2024} to replace algorithms such as spherical overdensity halo finders \texttt{AHF} \citep{Knollmann2009} and \texttt{ROCKSTAR} \citep{Behroozi2013}. Similar efforts also aim to directly learn the mapping between dark matter and galaxy fields through high-fidelity hydrodynamic simulations \citep{Villaescusa-Navarro2020a,Sether2024,Bourdin2024,Horowitz2025}. 

\begin{figure*}
    \centering
    \includegraphics[width=0.99\linewidth]{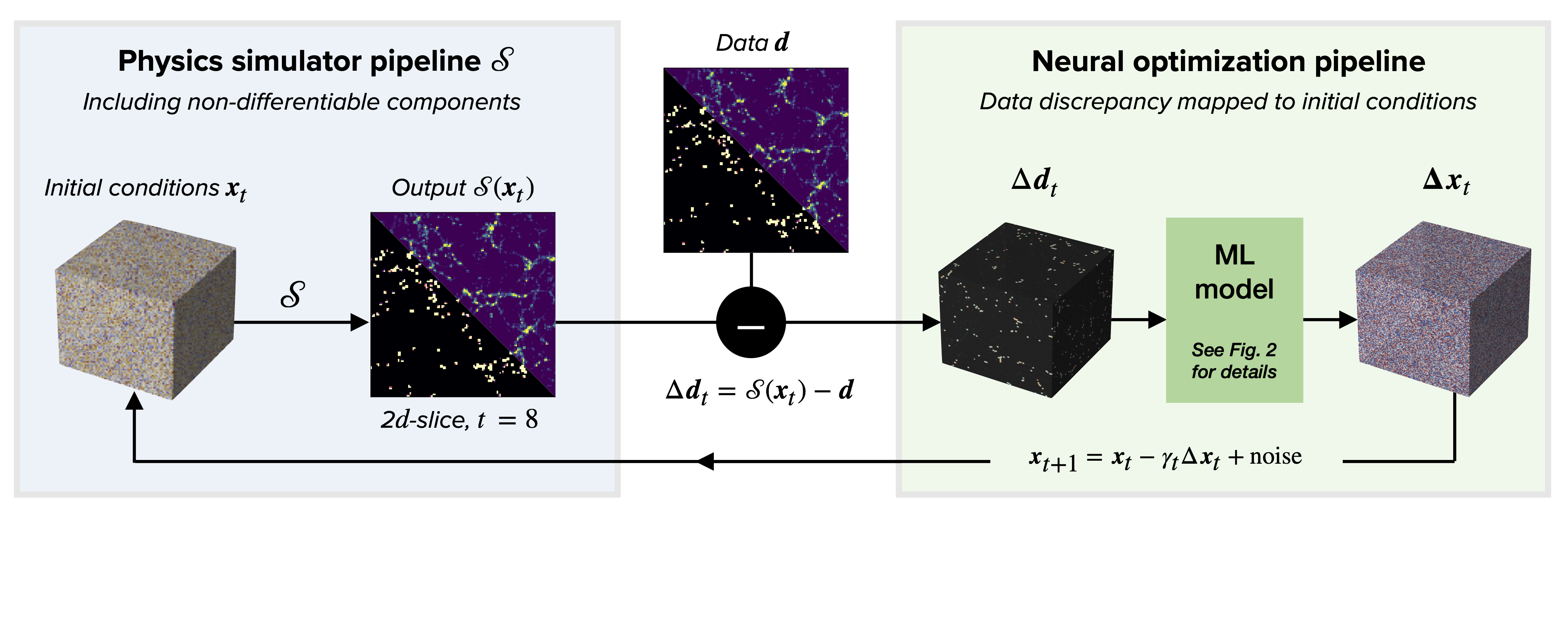}
    \vspace{-2.5em}
    \caption{High-level overview of \texttt{LULO} (Learning the Universe by Learning to Optimize) which aims to fit complex models to data by reconstructing the three-dimensional initial conditions. The process consists of two components: 1) applying a high-fidelity physics simulator $\mathscr{S}$, and 2) updating the initial conditions to minimize discrepancies $\Delta \ve{d}$ between the simulator output and the data. Importantly, any simulator model, including fully non-linear and non-differentiable ones, is supported. The neural optimizer, pre-trained via a supervised approach to learn how to map data discrepancies to updates in the initial conditions, proposes an update direction $\Delta \ve{x}$ across all initial condition voxels simultaneously. In the current implementation, the step size $\gamma_t$ is optimized via a line search algorithm that requires running the simulator (i.e., $\gamma_t=\gamma_t(\mathscr{S}, \Delta \mathbf{x}_t)$; see details in sections~\ref{sec:update_direction}). The iterative process continues until the simulation output aligns with the data, as shown in the $2d$-slices after eight optimization steps. The slices show the evolved non-linear dark matter density field in a cubic box with side length $250h^{-1}$ Mpc and the corresponding dark matter halo field as produced by a non-differentiable spherical overdensity algorithm.}
    \label{fig:L20_schematic}
\end{figure*}

To address these challenges of having to develop precise and accurate differentiable physics models, we propose a novel approach that optimizes cosmological data without requiring simulator differentiability. As strategy (ii) states, instead of accelerating the physical modelling through efficient approximations, we aim to accelerate the search process itself. This means that alternatives to full-scale simulations designed to reduce computational costs no longer need to be constrained by the requirement of differentiability, thereby establishing a new foundation for future model development. Our strategy is based on recent advances in \textit{learning to optimize} (L2O), as first proposed by \citet{Andrychowicz2016,Chen2016}, which uses machine learning to design and/or improve optimization methods \citep[also see reviews by][]{Chen2021a,Hospedales2022}. This so-called meta-learning or learning-to-learn, addresses the task of improving the learning algorithm itself. The machine-learning-based optimizer serves exclusively as a search engine, distinctly separated from the physics predictions, which are generated by state-of-the-art simulators, as illustrated in Fig.~\ref{fig:L20_schematic}. We thus address the common remark (see e.g \citet{Huertas-Company2022}) that physics knowledge should be in the physics models and not in the machine learning model, leading to higher interpretability and explainability. This also aligns with recent efforts in promoting the safe use of machine learning for research applications, as seen in e.g. \citet{Bartlett2024, Holzschuh2024}.

It is worth stressing that our approach differs conceptually from recent advances in generative modelling, including flow-based models and conditional diffusion models \citep{Legin2023,Holzschuh2024}. These aim to sample plausible initial conditions by learning the underlying distribution constrained by the data. This learning process relies on a training dataset consisting of an ensemble of initial and final condition pairs from simulations, a set also used in the training of field-level emulators of structure formation \citep{He2019,AlvesDeOliveira,Bernardini2020,Kaushal,Jamieson2022b, Jamieson2024}. In contrast to the aforementioned techniques, we aim to learn the complete dynamical behavior of the simulator by understanding how discrepancies $\Delta \ve{d}$ in data space relate to changes $\Delta \ve{x}$ in the initial conditions. In other words, while generative models and emulators use a neural network as a model (falling under strategy (i)), we use it as part of an optimization algorithm (strategy (ii)). As illustrated in Fig.~\ref{fig:L20_schematic}, our incremental approach then entails having the neural network propose an update direction $\Delta \ve{x}$ in the initial conditions space based on the current data prediction, progressively refining the initial condition reconstruction at each iteration. We highlight that unlike learned iterative reconstruction approaches \citep[see e.g.][]{Putzky2017,Adler2017} that have been explored in the context of large-scale structure analysis \citep{Modi2021}, our approach eliminates the need for the gradient of an explicit data likelihood term, enabling the use of arbitrarily complex data models without requiring further development cycles.

We name our algorithm \textit{Learning the Universe by Learning to Optimize} (\texttt{LULO}) and release this work as part of the Learning the Universe collaboration\footnote{\href{https://learning-the-universe.org/}{https://learning-the-universe.org/}}, which seeks to learn the cosmological parameters and initial conditions of the Universe by simultaneously leveraging machine learning, Bayesian forward modelling, and state-of-the-art simulations. The manuscript is structured as follows. In Section~\ref{sec:method}, we define the general problem and present our approach of using a neural optimizer to optimize black-box simulator functions, irrespective of differentiability. Section~\ref{sec:L20_cosmo} details the adaptation for cosmological initial conditions reconstructions, including training data design, choice of simulators with non-differentiable components, and training procedure for the neural optimizer. In section~\ref{sec:demonstration} we apply \texttt{LULO} and show the high quality of the reconstructed initial conditions. In section~\ref{sec:disc}, we further discuss the significance of non-linear modelling and data likelihoods, followed by a discussion on the computational cost and flexibility of \texttt{LULO} to incorporate arbitrary forward models. We end by presenting our conclusions in section~\ref{sec:conc}.

\section{Related work on learning to optimize}
\label{sec:DFO}

While many optimization algorithms exist, the no-free-lunch theorem \citep{Wolpert1997} states there is no universally best method for all objectives. Achieving improved performance requires task-specific algorithm design, which is time-intensive and demands significant manual effort to tune and validate pipelines, architectures, and hyperparameters. Learning to Optimize (L2O) has recently emerged as a promising approach to alleviate this challenge for complicated tasks by automating and accelerating the optimization procedure.

The foundation of L2O, as introduced by \citet{Li2016} and \citet{Andrychowicz2016}, is to replace traditional hand-crafted update rules with learned update rules, referred to as the \textit{optimizer}. Various approaches to using learned optimizers have been developed \citep{Chen2016, Li2017, Metz2020, Chen2021a, Zheng2022, Premont-Schwarz2022, Heaton2023}. These methods have, however, predominantly been applied to low-dimensional problems, whereas our task of inferring cosmological initial conditions involves jointly optimising millions of parameters of the primordial white noise field. In this work, we draw inspiration from these strategies to develop our high-dimensional \textit{Learning the Universe by Learning to Optimize} framework.

\section{Method: Gradient-free optimization}
\label{sec:method}
Our goal is to address the inverse problem of reconstructing the Universe's initial conditions from volumetric datasets, such as dark matter halo or galaxy count data. This task poses significant challenges due to the inclusion of complex, non-linear, and often non-differentiable cosmological simulations to causally connect the initial conditions with data. For this work, we explore handling non-differentiability by seeking a maximum likelihood estimate. 

After introducing the general inverse problem we want to solve in section~\ref{sec:problem}, we discuss in section~\ref{sec:ml_dfo} how to leverage machine learning to design a gradient-free optimizer capable of handling non-differentiable models and three-dimensional data. How to train such a neural optimizer to provide update directions in high-dimensional parameter spaces is then introduced in section~\ref{sec:update_direction}. 

\subsection{Problem Overview}
\label{sec:problem}
Our specific task is to learn a machine learning algorithm that can optimize a black-box function $f: \mathbb{R}^{n}\rightarrow\mathbb{R}$, given by $\ve{x} \mapsto f(\ve{x})$, where $\ve{x} \in \mathbb{R}^{n}$ is the input parameters. We assume that the function $f$ does not have a closed-form representation, is costly to evaluate, and does not allow the computation of gradients. This means we can only query the function at a given point $\ve{x}$ and obtain a response $y = f(\ve{x})$, but we have no information on its gradient or analytic form. 

For the purposes of this work, we can identify $f$ as the composition of two parts as
\begin{equation}
    f(\ve{x}) = \mathcal{L} \circ \mathscr{S} (\ve{x}) = \mathcal{L}(\mathscr{S} (\ve{x})),
\end{equation}
where $\mathscr{S}$ is an algorithm in the form of a simulation that translates between the input $\ve{x}$ and output $\mathscr{S}(\ve{x})$, and $\mathcal{L}$ is a likelihood evaluation between the output and the data $\ve
{d}$. 

We seek to find the maximum likelihood estimate $\ve{x}^*$ such that
\begin{equation}
    \ve{x}^*= \underset{\ve{x} \in \mathbb{R}^{n}}{\arg\max} f(\ve{x})
\end{equation}
with a limited number of function calls to $f$. The aim is to train an optimizer $h_{\mathrm{opt}}$ with parameters $w_{\mathrm{opt}}$, such that given a query point $\ve{x}_{t}$ at optimization step $t$ and corresponding query answers $y_{t}$, $h_{\mathrm{opt}}$ proposes the next query point $\ve{x}_{t+1}$ as
\begin{equation}
    \ve{x}_{t+1} = h_{\mathrm{opt}}(\ve{x}_{t}, y_{t}; w_{\mathrm{opt}}).
    \label{eq:sec_updates}
\end{equation}

\subsection{Incorporation of ML to learn black-box optimization}
\label{sec:ml_dfo}
In this work, we focus on a task categorized as \textit{learning to learn without gradient descent by gradient descent} \citep{Chen2016}, i.e. learning a parametric function such as a neural network \textit{by} gradient descent to learn how to optimize a black-box function $f(\ve{x})$. A key distinction to \citep{Chen2016} is that we bypass the requirement on back-propagation through $f(\ve{x})$ during the training phase, enabling our optimizer to be trained even when gradient information is not available. 

Gradient descent \citep[see e.g. review by][]{Ruder2016} is based on explicit differentiability as
\begin{equation}
\label{eq:update}
\ve{x}_{t+1} = \ve{x}_{t} - \gamma\, \nabla_{\ve{x}_{t}} f(\ve{x}_{t})
\end{equation}
where $\gamma \in \mathbb{R}_+$ is the learning rate. As $f(\ve{x_{t}}) = \mathcal{L}(\mathscr{S}(\ve{x_{t}}))$, the gradient $\nabla_{\ve{x}_{t}} f(\ve{x}_{t})$ is given by the chain rule, which involves differentiating through the simulator $\mathscr{S}$. Since $\mathscr{S}$ is non-differentiable, we seek to learn an alternative to this optimization scheme without relying on gradients. For the data discrepancy, we choose a voxel-wise mean-squared likelihood
\begin{equation}
    \mathcal{L}(\ve{x}_t) = \frac{1}{2}(\ve{d}-\mathscr{S}(\ve{x}_t))^\top\Sigma^{-1} (\ve{d}-\mathscr{S}(\ve{x}_t)),
    \label{eq:mse}
\end{equation}
with $\Sigma$ chosen to an identity covariance matrix. Based on Eq.~\eqref{eq:update}, the update step $\Delta \ve{x}_{t} = \ve{x}_{t+1} - \ve{x}_{t}$ for the $i$:th voxel can be expressed 
\begin{equation}
    [\Delta \ve{x}_{t}]_i \propto [\nabla_{\ve{x}_{t}} f(\ve{x}_{t})]_i = \sum_j \underbrace{\left(\ve{d}_j-[\mathscr{S}(\ve{x}_t)]_j\right)}_{[\Delta \ve{d}_t]_j} \frac{\partial [\mathscr{S}(\ve{x}_t)]_j}{\partial [\ve{x}_t]_i}.
    \label{eq:grad}
\end{equation}
As $\Delta \ve{d}_{t}$ can be computed, while the partial derivatives of $\mathscr{S}$ are not available, we introduce a neural network (NN) model to learn the mapping from the data discrepancy $\Delta \ve{d}_{t}$ to the update step $\Delta \ve{x}_{t}$ for the initial conditions, i.e., $\Delta \ve{x}_{t}=\textstyle \mathrm{NN}(\Delta \ve{d}_{t})$. This bypasses the need to differentiate through the simulator $\mathscr{S}$.
Experiments on including $\ve{x}_{t}$ as an additional input to the NN did not show any improvement, and we leave further investigation to future studies. We also note that there can be $n_d$ quantities in the data space that are informative about the initial state. Thus, in general, we have
\begin{equation}
    \Delta \ve{d}_{t} = [\Delta \ve{d}_{t}^1,\dots,\Delta \ve{d}_{t}^{n_d}].
    \label{eq:multi_data}
\end{equation}
We leverage the exploration-exploitation philosophy central to evolutionary computation techniques by incorporating two competing terms: one deterministic and one stochastic \citep{Holland1975}. Similar to e.g. \citet{Fornasier2021,Riedl2023} we introduce the latter by a new term $B(\ve{x}_{t})\mathcal{N}_{t}$ to Eq.~\eqref{eq:update} for some function $B(\ve{x}_{t})$ that vanishes as we approach the ground truth and a standard Gaussian random vector $\mathcal{N}_{t}$. We choose an $\ve{x}$-independent function $B(\ve{x}_{t}) = 0.1 \times 0.9^t$ and draw a new $\mathcal{N}_{t}$ at each $t$. While the deterministic term from the machine-learning optimizer guides the optimization towards promising regions, the stochastic term introduces a diffusion mechanism that injects randomness into the dynamics to increase the exploration across the search space. The combined update and query rule can now be defined as:
\begin{align}
    \ve{x}_{t+1} & = \ve{x}_{t} - \gamma_t \Delta \ve x_t = \ve{x}_{t} - \gamma_t [\mathrm{NN}(\Delta \ve{d}_{t}) + B(\ve{x}_{t})\mathcal{N}_{t}] 
    \label{eq:update_rule1} \\
    \Delta \ve{d}_{t+1} & = \ve{d} - \mathscr{S}(\ve{x}_{t+1}).
    \label{eq:update_rule2}
\end{align}
We next introduce how to determine the update direction $\mathrm{NN}(\Delta \ve{d}_{t})$ and the amplitude $\gamma_t$, which in the current implementation depends on the optimization step $t$ through $\Delta \ve x_t$. 

\subsection{Predicting update direction from data discrepancy}
\label{sec:update_direction}

Training our neural optimizer $\mathrm{NN}$ to learn the relationship between discrepancies in data space $\Delta \ve{d}$ to updates in the initial conditions $\Delta \ve{x}$ requires choosing an adequate neural architecture and loss function. 

\subsubsection{Neural architecture design}
We design the neural optimizer to a U-Net/V-Net architecture \citep{Ronneberger,Milletari2016}, similar to the one described in \citet{Jamieson2022b}, as these networks inherently capture fine spatial details in volumetric data. This is achieved through the use of convolutional operators working on multiple levels of resolution connected in a U-shape, first by several downsampling layers and then by the same amount of upsampling layers. The architecture is particularly suitable for accurately representing cosmological large-scale structures thanks to its sensitivity to information at different scales, which is maintained through the different levels as well as the skip connections between the down- and upsampling paths. To avoid information bottlenecks the number of filters were doubled during each downsampling step \citep{Szegedy2016,Ibtehaz2020}, resulting in a higher computational demand and increased performance. The convolutional network also naturally allows for multi-channel input, such as required when the data consists of several quantities as in Eq.~\eqref{eq:multi_data}. 

\subsubsection{Neural loss function}
To ensure convergence in iterative optimization, it is sufficient for updates to consistently move in the correct half-space, as this guarantees reaching the target for convex or quasi-convex problems (see Appendix~\ref{app:opt_guarantees}). We therefore choose a loss function that ensures that the predicted direction $\Delta \ve{x}_{\mathrm{pred}}=\mathrm{NN}(\Delta \ve{d})$ is maximally recovered. This is provided by the cosine similarity loss (CSL),
\begin{equation}
L_{\mathrm{csl}} = 1 - \cos(\Delta \ve{x}_{\mathrm{true}}, \Delta \ve{x}_{\mathrm{pred}}) = 1 - \frac{\Delta \ve{x}_{\mathrm{true}}^\top \Delta \ve{x}_{\mathrm{pred}}}{||\Delta \ve{x}_{\mathrm{true}}|| \cdot ||\Delta \ve{x}_{\mathrm{pred}}||},
\label{eq:cossim}
\end{equation}
which measures the similarity between the predicted and target vectors; a loss of zero would indicate perfect similarity. The choice of training data is also vital, but since it depends on the specific problem setup, we refer the reader to section~\ref{sec:training_data} for further details.

We have also explored a neural loss function based on mean squared error (MSE) alone and a hybrid loss combining MSE and CSL, but both alternatives resulted in poorer optimization performance. While MSE-based objectives ensure that the model effectively recovers the correct amplitude, they are prone to directional errors. Using the cosine similarity loss achieves a higher alignment, which is necessary for iterative optimization. 

The cosine similarity loss ensures accurate update directions but does not constrain the amplitude. To address this, we employ a systematic line search approach, iteratively adjusting the step size $\gamma_t \in \mathbb{R}$ along the predicted update direction $\Delta \ve{x}_t$. This process involves sequentially evaluating $\mathscr{S}(\ve{x}_t - \gamma_t\Delta \ve{x}_t)$, approximately four times for our set-up, to minimize the discrepancy to the data $\ve{d}$. We provide more details on the line search method in Appendix~\ref{app:line_search}. 

\section{Learning the Universe by Learning to Optimize (\texttt{LULO})}
\label{sec:L20_cosmo}

Our optimization strategy now needs to be adapted for the task of reconstructing cosmological initial conditions. This involves training the neural optimizer using a problem-specific dataset $\{\Delta \ve{x}, \Delta \ve{d}\}$ with $\ve{x}\mapsto \mathscr{S}(\ve{x})=\ve{d}$, where $\ve{x}$ represents the cosmic initial conditions as a white noise field, $\mathscr{S}$ is a black-box physics simulator that may not be differentiable, and $\ve{d}$ refers to the data, which can be either simulated or observed. Throughout this work, we use cubic three-dimensional fields of $128^3$ voxels both for the initial conditions and fields in data space. We discuss the process of generating the training data in section~\ref{sec:training_data}, the simulator model in section~\ref{sec:cosmo_sims}, and the specifics of the neural training procedure, including its performance, in section~\ref{sec:neural_training}.

\subsection{Training data generation}
\label{sec:training_data}

We extend the approach of \citet{Sarafian2020} who samples pairs of inputs and outputs of a black-box function and then averages the numerical directional derivatives over small volumes to obtain the mean-gradient of the true gradient. Our high-dimensional input consisting of $128^3 \sim 2\times10^6$ white noise field voxels does, however, not allow for perturbing one input parameter at the time. Instead, we simultaneously and stochastically perturb all voxels.

\subsubsection{Perturbing cosmological initial conditions}

\begin{figure*}
    \centering
    \includegraphics[width=1.0\linewidth]{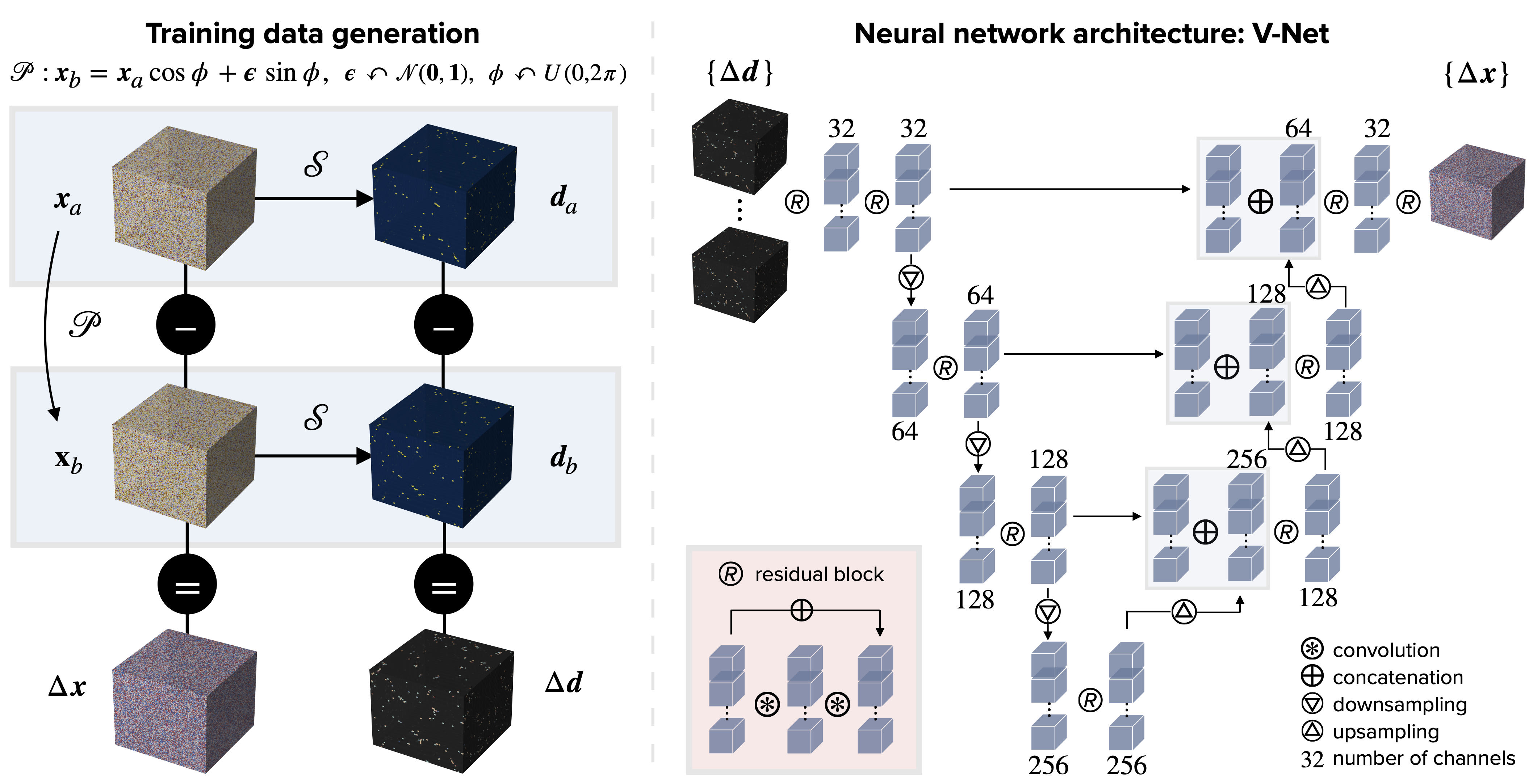}
    \caption{Learning the relationship between changes in the output of a cosmological simulator and corresponding changes in the initial conditions involves two key steps: generating training data (left) and training a machine learning model with a specific architecture (right). We perturb the initial conditions using an operator $\mathcal{P}$ that preserves the statistical properties of the Gaussian initial conditions, namely zero mean and unit variance. The resulting set of difference fields $\{\Delta \ve{d}\}$, where $\Delta \ve{d} = \ve{d}_a - \ve{d}_b$, serves as input to the neural optimizer model during training. The neural optimizer is tasked with predicting the corresponding update, $\Delta \ve{x} = \ve{x}_a - \ve{x}_b$, in the initial conditions. The chosen architecture is a convolutional V-Net model, which effectively handles structures and their correlations across multiple scales. To prioritize learning accurate update directions we use a cosine similarity loss during training.}
    \label{fig:training_data_schematic}
\end{figure*}

Let us assume $\ve{x}_a$ is a particular vector representing the three-dimensional initial white noise field of a cosmological volume. In our cosmological setting, it is drawn from a multivariate normal Gaussian with zero mean and diagonal unit co-variance. The pair of initial and final conditions ($\ve{x}_a,\ve{d}_a$) is created through our physics simulator $\mathscr{S}: \ve{x}_a \rightarrow \ve{d}_a$. Now we create a second pair, ($\ve{x}_b,\ve{d}_b$), by perturbing the initial conditions as
\begin{equation}
    \ve{x}_b = \ve{x}_a \cos{\phi}  + \ve{\epsilon} \sin{\phi}  \, ,
    \label{eq:perturbing_ics}
\end{equation}
where the phase variable $\phi \curvearrowleft U(0,b)$ is drawn from a uniform random distribution and  $\ve{\epsilon} \curvearrowleft \mathcal{N}(\ve{0},\mathds{1})$ is another random Gaussian white noise field. This procedure ensures that the perturbed $\ve{x}_b$ has the same unit variance as $\ve{x}_a$. We choose $b=2\pi$ which allows for $\ve{x}_b$ to not only be a perturbation to $\ve{x}_a$, but possibly a completely new or even a fully anti-correlated realization. We then evaluate the corresponding simulation at $\ve{x}_b$ and obtain the corresponding predicted data $\ve{d}_b$. The vector differences $\Delta \ve{x}$ and $\Delta \ve{d}$ can now be formulated as
\begin{align}
    \Delta \ve{x} & = \ve{x}_a -\ve{x}_b\\ 
    \Delta \ve{d} & = \ve{d}_a-\ve{d}_b
\end{align}
We visualize this training generation process in Fig.~\ref{fig:training_data_schematic}. We construct the training ensemble by randomly creating $\ve{x}_a$ and $\ve{x}_b$ and repeating the above procedure. In this fashion, the set of $\{\Delta \ve{d}, \Delta \ve{x}\}$ pairs provide the training data, with $\Delta \ve{d}$ being the input and $\Delta \ve{x}$ the output. The network is then trained in a supervised fashion using the cosine similarity loss function (Eq.~\eqref{eq:cossim}). 

\subsection{Non-linear, non-differentiable cosmological simulator}
\label{sec:cosmo_sims}
To translate between the initial conditions $\ve{x}$ and the data space, we employ a high-fidelity $N$-body simulation coupled with a non-differentiable spherical halo finder. The simulations are performed with $128^3$ particles within a cubic volume with a side length of $250h^{-1}$ Mpc, i.e. at a Lagrangian resolution of $1.95h^{-1}$ Mpc. The resulting particles and haloes are discretized onto grids of $128^3$ elements using a cloud-in-cell mass assignment scheme to obtain the density contrasts. The dark matter overdensity and halo overdensity thus maintain a resolution of $1.95h^{-1}$ Mpc, which enables accurate representation of non-linear cosmic structures. The number density of haloes, used as tracers in our mock setup, is $\sim 2.2 \times 10^{-4}$ $(h/\mathrm{Mpc})^3$, which is comparable to the tracer number density in galaxy surveys \citep[see e.g.][]{Ata_2017, euclidcollaboration2024euclidiovervieweuclid}.

\subsubsection{$N$-body simulation}
$N$-body simulations are currently the most advanced numerical technique for modeling the full non-linear structure formation of the Universe \citep{Vogelsberger2020,Angulo2021}. In this work, we use the $N$-body simulation code \texttt{Gadget-IV} \citep{Springel2021}. From the white-noise field, the simulator initial conditions of particles are generated at $z=127$ using \texttt{MUSIC} \citep{Hahn2011} together with transfer function from Eisenstein \& Hu \citep{Eisenstein1998,Eisenstein1999}. The cosmological parameters used are $\Omega_\mathrm{m} = 0.3175$, $\Omega_\mathrm{b} = 0.049$, $h = 0.6711$, $n_\mathrm{s} = 0.9624$, $\sigma_8 = 0.834$, and $w=-1$ consistent with latest constraints by Planck \citep{Aghanim2018} and equivalent with the fiducial cosmology of the Quijote simulations \citep{Villaescusa-Navarro2020}. We compile and run the dark-matter-only and TreePM code version of \texttt{Gadget-IV} with $\mathrm{PMGRID}=256$. Although in theory differentiable, achieving this would require storing the full phase-space information at every time step of the simulator integrator—on the order of thousands—which memory-wise is infeasible.

\subsubsection{Dark matter haloes}
\label{sec:amiga}
To identify haloes in our $N$-body simulations, we use the spherical-overdensity-based algorithm \citep{Press1974,Warren1992,Lacey1994} as implemented by the Amiga Halo Finder (\texttt{AHF}) \citep{Gill2004,Knollmann2009}. We require at least $20$ particles per halo and adopt the \( M_{200\mathrm{c}} \) mass definition, which represents the mass enclosed within a spherical region where the average density is $200$ times the critical density. This definition is particularly stringent, as it necessitates precise resolution of the density profile down to small scales. We generate haloes in $10$ mass bins logarithmically separated between $1\times 10^{13}M_{\odot}h^{-1}$ and $2\times 10^{15}M_{\odot}h^{-1}$, which for our particle mass of $6.57 \times 10^{11} M_{\odot}h^{-1}$ ensures all found haloes are incorporated. We emphasize that \texttt{AHF} uses a non-differentiable algorithm to find haloes from the positions and velocities of simulated particles.

For the haloes, we want to retain some mass information during the assignment, as otherwise a low mass halo would be regarded as significant as a massive halo. We do this by applying a mass weighting to the final halo overdensity field $\boldsymbol \delta^{\mathrm{halo}}$ as
\begin{equation}
    \boldsymbol \delta^{\mathrm{halo}} = \frac{\sum_iN_if(M_i)\boldsymbol \delta_i}{\sum_i N_if(M_i)},
    \label{eq:overdensity_weighting}
\end{equation}
with $N_i$ being the number of haloes, $f(M_i)$ a function of the average mass of haloes, and $\boldsymbol \delta^{\mathrm{halo}}_i$ the over-density field in halo mass bin $i$. In this work, we choose to make use of two distinct mass functions. The first one is a mass-weighted field through
\begin{equation}
    f(M_i) = \frac{M_i}{1+\sqrt{M_i}}
    \label{eq:mass_weighting}
\end{equation}
with $M_i$ expressed in units of $10^{14}h^{-1}M_{\odot}$. The denominator in the mass-weighting was introduced by \citet{Seljak2009} to down-weight the most massive haloes, as their contribution is otherwise too dominant. While more optimal weighting schemes have been proposed \citep{Hamaus2010, Liu2021}, we adopt Eq.~\eqref{eq:mass_weighting} and leave further exploration for future studies. The second function we adopt is a halo count overdensity, defined by $f(M_i) = 1$ in Eq.~\eqref{eq:overdensity_weighting}, which is equivalent to directly creating an overdensity field from all haloes in one single wide mass bin. We choose the data to be these two halo fields
\begin{equation}
    \ve{d} = [\boldsymbol{\delta}^{\mathrm{halo}}_{\mathrm{mw}}, \boldsymbol{\delta}^{\mathrm{halo}}_{\mathrm{count}}],
\end{equation}
where $\boldsymbol{\delta}^{\mathrm{halo}}_{\mathrm{mw}}$ is the mass-weighted (mw) halo count overdensity field, and $\boldsymbol{\delta}^{\mathrm{halo}}_{\mathrm{count}}$ is the halo count overdensity.

\subsection{Neural training procedure}
\label{sec:neural_training}

In Fig.~\ref{fig:training_data_schematic}, we present a schematic overview of the training process for the neural optimizer. After generating the training data via perturbations with the cosmological simulator $\mathscr{S}$, the resulting pairs of difference fields are used as input (${\Delta \ve{d}}$) and output ($\Delta \ve{x}$) for supervised training of the neural network. For details on the computational cost of training data generation and neural network training, we refer the reader to section~\ref{sec:compcost}.

\subsubsection{Size of training data}
We generate $1000$ random initial conditions, run the simulation pipeline, and use Eq.~\eqref{eq:perturbing_ics} to perturb each initial condition $5$ times. For each of the $6000$ simulations, we store the initial white noise field $\ve{x}$, the redshift $z=0$ snapshot from \texttt{GADGET-IV} as well as the halo catalog provided by \texttt{AHF}. In total, we create $5000$ difference fields for the initial conditions $\Delta \ve{x}$ as well as for the data $\Delta \ve{d} = [\Delta \boldsymbol{\delta}^{\mathrm{halo}}_{\mathrm{mw}}, \Delta \boldsymbol{\delta}^{\mathrm{halo}}_{\mathrm{count}}]$. 

During training, we further augment the training data set through rotations and flips, similar to the training procedures in \citet{He2019,Jamieson2022b}. Additional combinations of simulations from the set of $5000$ could have been generated. Tests of incorporating such pairs showed no improvement, suggesting that these fully uncorrelated fields do not add any additional information. 

\subsubsection{Training set-up}
We use the framework \texttt{map2map}\footnote{\href{https://github.com/eelregit/map2map}{github.com/eelregit/map2map}} for field-to-field emulators, based on \texttt{PyTorch} \citep{Paszke2019}, to train the neural network. We allocate most simulation pairs to training, reserving only $2\%$ ($100$ random pairs) for validation to monitor the cosine similarity loss and prevent overfitting. As each application of the trained neural optimizer in \texttt{LULO} will be performed on new initial conditions distinct from the ones used during training (hence effectively being the test set), we only split the data into a training and a validation set.  

Parallel processing of large volumes into sub-boxes is enabled. In particular, we crop the $128^3$ fields to $64^3$ due to memory requirements and use a batch size of $16$ over multiple \texttt{NVIDIA} A$100$ $40$GB Tensor Core GPUs. We preserve translation symmetry by periodically padding each sub-box by $48$ voxels on each side. Additionally, we randomly reflect and/or apply rotations of the sub-boxes in multiples of $90$ degrees, thus covering a discrete set of orientations rather than all possible rotations, resulting in rotational symmetry being approximately preserved. We use an initial learning rate of $4 \times 10^{-4}$ with the AdamW optimiser \citep{Loshchilov2019}, which is an extension to the Adam optimiser \citep{Kingma2014} using decoupled weight decay. We use a decay coefficient $\lambda=1\times 10^{-4}$ and parameters $\beta_1=0.9$ and $\beta_2=0.99$. At a plateau, if the loss is not reduced by a factor $0.999$ over $3$ epochs, we half the learning rate. To achieve high performance, we primarily monitor the cosine similarity loss for the training and validation sets, but the performance after incorporation within the full optimization process is also evaluated.



\subsubsection{Performance}
As an initial assessment of the trained model's performance, we evaluate it on the validation set, while acknowledging that the true evaluation will occur on the \textit{test set}, that is when the neural optimizer is applied in \texttt{LULO} to reconstruct the initial conditions. The cosine similarity ($1-\mathrm{loss}$) achieved on the validation set is $0.087 \pm 0.024$. Applying a Gaussian smoothing filter with standard deviation $10h^{-1}$ Mpc to both the predicted $\Delta \ve{x}_{\mathrm{pred}}$ and the true $\Delta \ve{x}_{\mathrm{true}}$, reveals a higher alignment of cosine similarity $0.944 \pm 0.036$ at these larger scales. 

\begin{figure}
    \centering
    \includegraphics[width=1.0\linewidth]{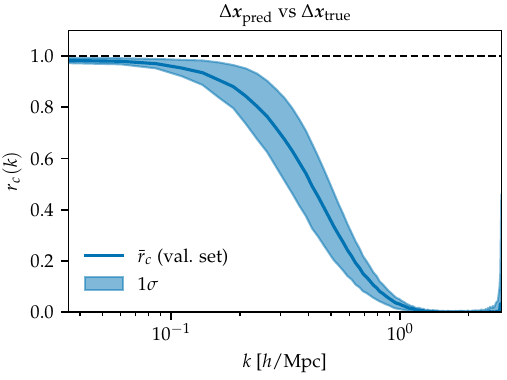}
    \vspace{-1.5em}
    \caption{The performance of the neural optimizer is evaluated through the cross-correlation between the predicted update direction $\Delta \ve{x}_{\mathrm{pred}}$ in the initial conditions and the ground truth $\Delta \ve{x}_{\mathrm{true}}$. The mean and standard deviation of the cross-correlation across all samples in the validation set are shown. A high alignment between the prediction and the truth down to scales of approximately $0.3h$ Mpc$^{-1}$ is obtained, with the correlation decreasing at smaller scales. This provides insight into which scales in the initial conditions are crucial for correcting data discrepancies at our particular resolution.} \label{fig:optimizer_performance}
\end{figure}

In Figure~\ref{fig:optimizer_performance} we quantify the output of the trained model across all scales using the cross-correlation between the predicted $\Delta \ve{x}_{\mathrm{pred}}$ and the true $\Delta \ve{x}_{\mathrm{true}}$. The higher correlation observed at larger scales, with a decrease at smaller scales, can be attributed to the fixed data resolution and the hierarchical nature of structure formation. Specifically, due to gravitational collapse, larger regions in the initial conditions evolve into smaller structures in the data \citep{Gunn1972}. As the resolution of the initial conditions matches that of the data in our setup, small-scale structures in the initial conditions collapse below the data resolution, where information cannot be accessed, as also discussed in e.g. \citet{Doeser2024}. This coupling also means that to explain non-linear scales in the data, linear scales in the initial conditions need to be accurately recovered. 

\begin{figure*}
    \centering
    \includegraphics[width=1\linewidth]{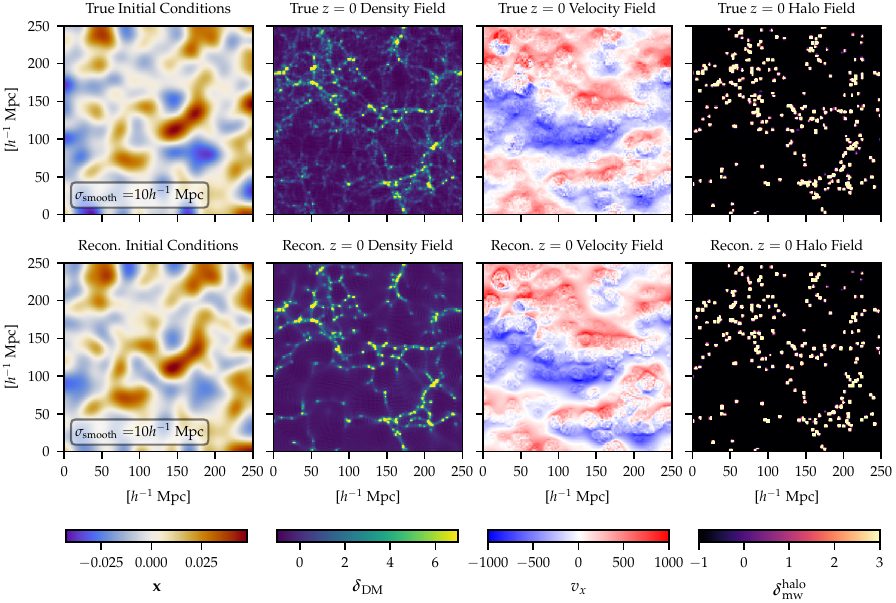}

    \caption{The reconstructed initial conditions (bottom row; left), and the corresponding forward simulation output at $z=0$ in terms of dark matter overdensity (centre left), the Lagrangian velocity field (centre right), and mass-weighted halo count overdensity field (right) are compared with the ground truth (top row). The initial conditions have been smoothed at $10h^{-1}$ Mpc to reveal the visual alignment. A high correlation over these scales suffices for a highly accurate reconstruction in the data space, even at the smallest non-linear scales. In particular, the density and velocity fields, despite only being by-products and not used to constrain the initial conditions, are well reconstructed. Note that for the mass-weighted field, all haloes as found by the non-differentiable \texttt{AHF} algorithm have been included. Low-mass haloes, being the most numerous, are often not accurately reproduced because of the low number of particles per halo (see section~\ref{sec:halo_reconstruction}). The majority of haloes are nonetheless observed to be accurately recovered.}
\label{fig:reconstruction_viz}
\end{figure*}

\section{Initial conditions reconstruction with \texttt{LULO}}
\label{sec:demonstration}

To test \texttt{LULO} we start by generating a $M_{200\mathrm{c}}$ dark matter halo catalogue using the forward simulation pipeline given in section~\ref{sec:cosmo_sims}. We emphasize the non-linear and non-differentiable nature of our setup, which employs $128^3$ particles within a cubic volume spanning $250h^{-1}$ Mpc per side. Throughout the optimization process, we consistently use the same pipeline and use $128^3$ voxels to grid density, velocity, and halo fields. The ground truth initial conditions are not part of the training or the validation data sets, ensuring the neural optimiser's generalizability is tested. From the halo catalogue, we create a mass-weighted halo count overdensity and a halo count overdensity that together makes the data $\mathbf{d} = [\boldsymbol \delta^{\mathrm{true}}_{\mathrm{mw}},\boldsymbol \delta^{\mathrm{true}}_{\mathrm{count}}]$. 

After describing the initialization of the optimization process in section~\ref{sec:towards_optimum}, we showcase the high accuracy of the reconstructed initial conditions and the late-time dark-matter and halo fields in section~\ref{sec:reconstructions}. While most of this section focuses on quantifying the reconstruction quality using a single set of ground truth initial conditions and corresponding data, section~\ref{sec:various_data} presents additional examples of reconstructions with other ground truths.

\subsection{Initialization}
\label{sec:towards_optimum}
The initial guess for the initial conditions is chosen as a zero field. Given the high-dimensional space of $128^3 \sim 10^6$ voxels and the Gaussian nature of the true initial conditions, characterized by a zero mean and unit variance, a zero field is closer to the truth than a randomly generated white noise field. Due to the reduced initial cosmological power, the first simulation yields no haloes. This means the first input to the neural optimizer is $\Delta \ve{d} = \ve{d}$, from which the first search direction $\Delta \ve{x}$ is predicted. The first mapping initiates the optimization process of iteratively updating the initial conditions per Eqs.~\eqref{eq:update_rule1}–\eqref{eq:update_rule2} such that they, after forward simulation, increasingly match the data. The full monitoring of the subsequent optimization progress and iterative improvements is presented in Appendix~\ref{app:towards_opt}. 

\subsection{Reconstructed initial conditions}
\label{sec:reconstructions}
\begin{figure*}
    \centering
    \includegraphics[width=1.0\linewidth]{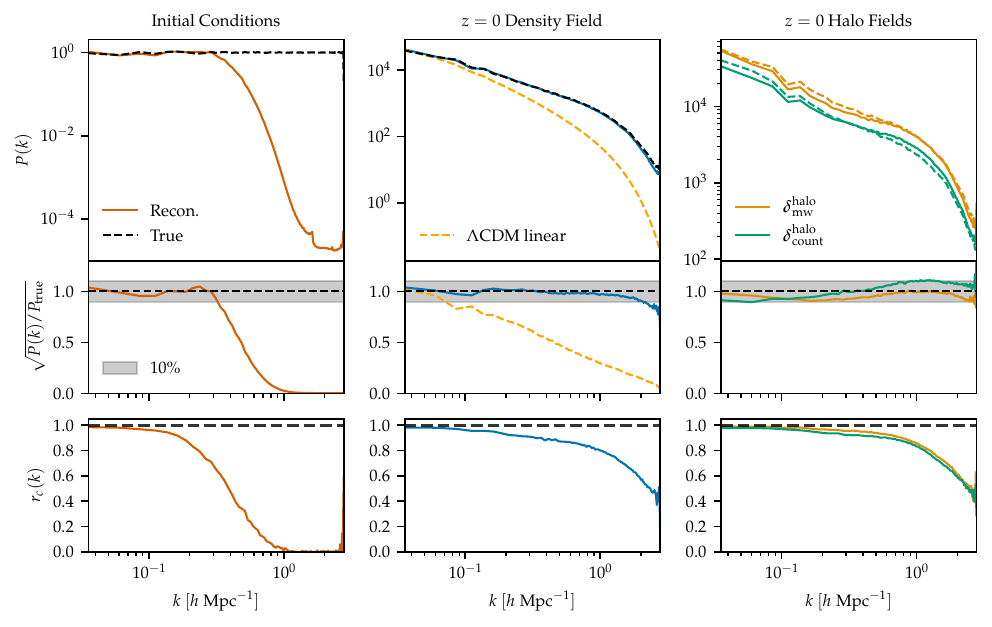}
    \vspace{-1.8em}
    \caption{The quality of the reconstruction is quantified through the power spectra (top) and cross-correlation (bottom) of the reconstructed initial conditions (left), the dark matter overdensity (center), the mass-weighted halo count overdensity and halo count overdensity (right). In the residual panels, we display the transfer function, i.e. the square root of the ratio of the power spectra of the reconstructions with the ground truth. With initial conditions matching to within $10\%$ down to scales of $\sim 0.3h$ Mpc$^{-1}$, the final density field and the halo fields match down to highly non-linear scales of $\sim 2.1h$ Mpc$^{-1}$ and $\sim 2.7h$ Mpc$^{-1}$, respectively. The cross-correlations further demonstrate that high correlation in the initial conditions at large scales ($80\%$ at $\sim0.23h$ Mpc$^{-1}$) is sufficient for an $80\%$ correlation in the dark-matter overdensity and halo fields beyond scales of $1h$ Mpc$^{-1}$. This reflects the gravitational collapse of proto-structures into final structures as discussed in section~\ref{sec:reconstruction_powspec_etc}. The linear $\Lambda$CDM power spectrum at $z=0$ is shown as a reference to highlight the non-linearities involved.}
    \label{fig:reconstruciton_powspec}
\end{figure*}

In Fig.\ref{fig:reconstruction_viz}, we visualize the reconstruction after $22$ optimization steps, equivalent to running the full simulation pipeline $96$ times, by which point the data discrepancy has converged (as shown in Appendix~\ref{app:towards_opt}). We show the initial conditions and the corresponding derived fields at redshift $z=0$, including the dark-matter overdensity field, one component of the Lagrangian velocity field (i.e., particles are gridded as their ordering in the initial conditions such that each voxel corresponds to a particle velocity), and the mass-weighted halo count overdensity. To visualize the correlation in the initial conditions we smooth the field over $10h^{-1}$ Mpc with a Gaussian kernel. 

The sole objective of the optimizer is to update the initial conditions such that the data discrepancy in the halo fields is reduced. In the mass-weighted halo count overdensity, we see that most haloes are accurately reconstructed in the correct regions. The underlying density and velocity fields, arising as by-products, are also well reconstructed. We note that in the dark matter overdensity field, regions of high density are reconstructed with greater accuracy. This outcome is expected, as these regions correspond to the more massive haloes, which provide stronger constraints. The Lagrangian velocity field, while presenting a particularly stringent test as individual simulated particle velocities are compared, demonstrates notable alignment, capturing large-scale coherent flows in the correct regions. 

\subsubsection{Reconstruction accuracy and temporal mode coupling}
\label{sec:reconstruction_powspec_etc}

Next, we quantitatively compare the reconstructed initial conditions and the corresponding evolved fields with their ground truths. We quantify the reconstruction through the spatial correlation of structures in terms of the power spectra and cross-correlation. We make use of the \texttt{PYLIANS3}\footnote{\href{https://github.com/franciscovillaescusa/Pylians3}{https://github.com/franciscovillaescusa/Pylians3}} package \citep{Pylians} to compute these quantities for the initial conditions, the dark matter overdensity, the mass-weighted halo overdensity and the halo overdensity. We show the result in Fig.~\ref{fig:reconstruciton_powspec} and also include the transfer function (the square root of the ratio of the power spectra of the reconstructions with the ground truth). For comparison and to highlight the non-linear scales involved in this reconstruction, we display the linear $z=0$ power spectrum computed with \texttt{CLASS} \citep{Blas2011}.

The transfer function of the initial conditions matches to within $10\%$ for scales $k \leq 0.33h$ Mpc$^{-1}$, after which it drops. The final density field still matches within $10\%$ to highly non-linear scales of $k = 2.09h$ Mpc$^{-1}$. Similarly, the halo fields stay within $12\%$ for all scales $k \leq 2.69h$ Mpc$^{-1}$. For comparison, the Nyquist frequency of the box is $k_N = 2.79h$ Mpc$^{-1}$, indicating that the agreement extends close to the resolution limit. We note that to explain the data at non-linear scales and obtain $80\%$ correlation in the dark-matter overdensity field, mass-weighted halo overdensity and halo overdensity to scales of $1h$ Mpc$^{-1}$, $1.28h$ Mpc$^{-1}$, and $1.15h$ Mpc$^{-1}$, respectively, the method requires high cross-correlations in the initial conditions at large scales ($80\%$ for $k\leq0.23h$ Mpc$^{-1}$). This also connects back to the cross-correlation in Fig.~\ref{fig:optimizer_performance} for $\Delta \ve{x}_{\mathrm{pred}}$ vs $\Delta \ve{x}_{\mathrm{true}}$, where we identified that linear scales in the initial conditions need to be updated to minimize the data discrepancy at non-linear scales. To highlight the improvement achieved by modeling non-linear scales, we compare our reconstruction to the optimal linear reconstruction in Appendix~\ref{app:linear_recon}.

This coupling of scales naturally follows from hierarchical growth of structures, with primordial objects collapsing through gravitational interactions into smaller regions \citep{Gunn1972}. The amount of information in the initial conditions at the field-level required to account for highly non-linear structures observed today was also discussed in \citet{Doeser2024}, where it was demonstrated that modelling of late-time dark-matter structures to non-linear scales of $2.79h$ Mpc$^{-1}$ is needed to recover initial conditions with a cross-correlation of $80\%$ to scales of approximately $0.35h$ Mpc$^{-1}$.

The transfer function and cross-correlation furthermore show that the mass-weighted halo overdensity is slightly more accurately recovered than the halo overdensity. This is to be expected, as the halo overdensity treats all haloes equally regardless of their mass. In contrast, the mass-weighted halo field incorporates mass information by assigning greater weight to more massive haloes through Eq.~\eqref{eq:mass_weighting}, making this halo field more informative and thus constraining. 

\subsubsection{Peculiar velocities of simulated particles}
By reconstructing the initial conditions, we gain access to the complete dynamical evolution of structures, including their velocities at non-linear scales. As \citet{Stiskalek2025} has recently demonstrated, field-level inference of initial conditions using non-linear modelling yields more accurate reconstructions of the peculiar velocity field than alternative methods. Velocities are thus a valuable by-product of our method and provide a critical tool for testing and differentiating cosmological models, including for example the growth rate of the large-scale structure \citep[see e.g.][]{Koda2014,Howlett2017,Boruah2020}. Reconstructions of non-linear peculiar velocities can further enhance the precision of such tests and are also vital for peculiar velocity corrections in local measurements of the expansion rate of the Universe, i.e. the Hubble constant \citep[e.g.][]{Mukherjee2021,Boruah2021,Kenworthy2022,Riess2022,Peterson2022,Carreres2025}. The impact of peculiar velocities on the measurement of $H_0$ from gravitational waves and megamasers was studied in \citet{Boruah2021}. 

\begin{figure}
    \centering
    \includegraphics[width=1.0\linewidth]{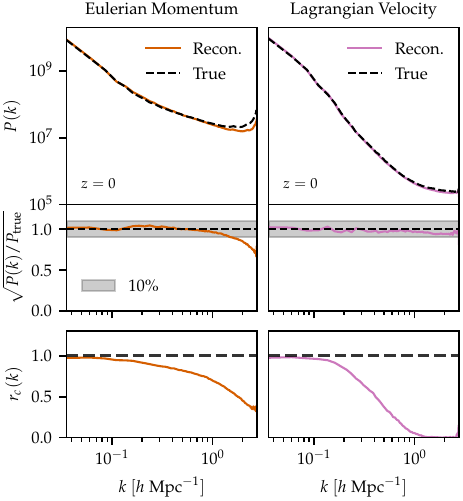}
    
    \caption{The reconstructed $z=0$ momenta and velocities of the individual simulation particles are quantified through the power spectra (top) and cross-correlation (bottom). For both the Eulerian momentum and Lagrangian velocity field, the power spectra and cross-correlations of the individual spatial components have been averaged over. The power spectra show high accuracy as displayed by the transfer function well into the non-linear regime. }
    \label{fig:dm_vel}
\end{figure}

In Fig.~\ref{fig:dm_vel}, we show the power spectra, transfer function and cross-correlation of the Eulerian momentum field and the Lagrangian velocity field. For both fields, we average the power spectra and the cross-correlation over the three spatial components of the velocities. Since each particle in the simulation has the same mass, to obtain the momentum field we weigh the particle positions by their velocity when gridding the particles. For the Lagrangian velocity field, we order the particles according to their position in the $3$D grid of the initial conditions and store their velocity in each component. 

In both the Eulerian and Lagrangian case, we achieve a high level of accuracy, with transfer functions within $10\%$ down to $k=1.9h$ Mpc$^{-1}$ and $k=2.7h$ Mpc$^{-1}$, respectively. The cross-correlation is $80\%$ down to scales of $k=0.59h$ Mpc$^{-1}$ and $k=0.24h$ Mpc$^{-1}$, respectively. The Eulerian momentum is expected to show a higher cross-correlation, as it also depends on the positions of the particles. In contrast, the Lagrangian velocity field only involves the velocities of individual simulated particles, which is more challenging to reconstruct.

\begin{figure}
    \centering
    \includegraphics[width=1.0\linewidth]{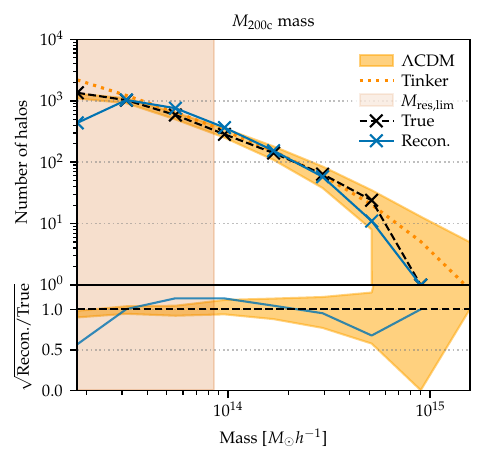}
    \vspace{-1em}
    \caption{The halo mass function shows that most of the $M_{200\mathrm{c}}$ halo masses are accurately recovered, with a slight over-estimation in the number of low-intermediate mass haloes and an under-estimation in the intermediate-high mass haloes. The mass resolution limit at $130$ particles, below which the mass is less accurately resolved, is displayed in the shaded region. Above the limit, the reconstruction is consistent with $\Lambda$CDM (yellow) within $3\sigma$, as derived from the $1000$ independent simulations in the training data. The residual panel shows the square root of the ratio between the reconstruction and the ground truth, and between the $\Lambda$CDM band and the truth.}
    \label{fig:reconstruction_hmf}
\end{figure}

\begin{figure}
    \centering
    \includegraphics[width=1.0\linewidth]{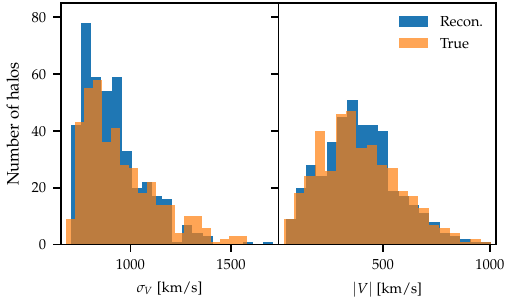}
    \caption{The reconstructed distributions of the halo velocity dispersion and the halo peculiar velocity of all haloes above the mass resolution limit. Velocity information was not included in the data, but the reconstructions in these properties still show a high degree of accuracy.}
    \label{fig:halo_vel}
\end{figure}

\begin{figure*}
    \centering
    \includegraphics[width=1.0\linewidth]{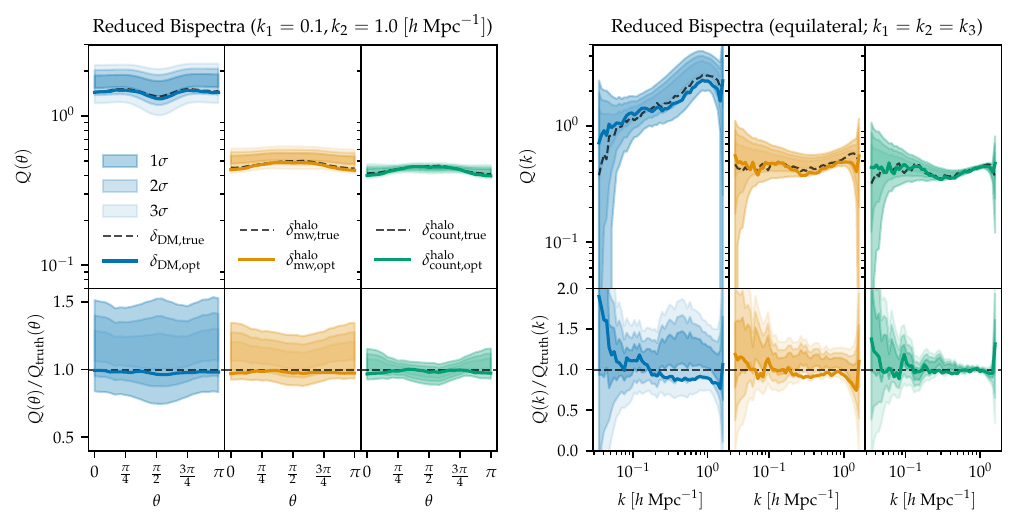}
    \vspace{-1em}
    \caption{The reduced bispectra in two distinct triangular configurations are used to assess the reconstruction accuracy: a configuration defined by fixed wave vector magnitudes as a function of the separation angle $\theta$ (left) and an equilateral triangle configuration with equal wave vector magnitudes to examine all scales (right). The three panels display the results for the final density field, the mass-weighted halo overdensity, and the halo overdensity, along with residuals comparing the reconstructed fields to the ground truth. We also show cosmic variance ($1\sigma$, $2\sigma$, and $3\sigma$ for $\Lambda$CDM) derived from the $1000$ unperturbed, and hence uncorrelated, training simulations (shaded regions). In particular, we note that the largest scales are dominated by cosmic variance. In all six bispectra shown, the reconstructions match the underlying theory and ground truth to within $5\%$ (left) and $10\%$ (right) over most angles and scales.}
    \label{fig:recon_bispec}
\end{figure*}

\subsubsection{Halo mass function}
\label{sec:halo_reconstruction}
The halo mass distribution, i.e. the number of haloes as a function of mass, serves as an important probe for testing the underlying cosmological model \citep[see, e.g.,][]{Jenkins2001a,Artis2021a,Stopyra2021}. Despite our method operating within a non-differentiable structure formation framework, our reconstruction demonstrates accurate recovery across most mass ranges. In Fig.~\ref{fig:reconstruction_hmf}, we show the halo mass function of the reconstruction compared with the ground truth and the theoretical $\Lambda$CDM prediction using \texttt{Tinker} \citep{Tinker}. We also leverage the $1000$ independent simulations from the training set to estimate the scatter within $\Lambda$CDM in each mass bin. In the study by \citet{Mansfield2020}, it was shown that at least $N=130$ particles are required to accurately describe the density profile and mass of $M_{200\mathrm{c}}$ dark matter haloes. For our chosen set of cosmological parameters and number of simulated particles, this corresponds to $8.53\times 10^{13}h^{-1} M_{\odot}$. We show this mass resolution limit with the red shaded region in Fig.~\ref{fig:reconstruction_hmf}. While this limit is crucial for recovering accurate density profiles, we note that it is a conservative choice for obtaining an accurate halo mass function. In our case, deviations from the theoretical prediction only begin at $\sim 4 \times 10^{13}h^{-1} M_{\odot}$. It is important to note that this arises from the current simulation setup and that our method can also be applied to higher-resolution simulations (see section~\ref{sec:compcost}).

Our reconstructed halo catalogue aligns well with the theoretical prediction and only deviates from the ground truth near $5\times 10^{14}h^{-1}M_{\odot}$. In total, we identify $2823$ haloes in the reconstructed field, with $432$ above the mass resolution limit. While this total is lower than the $3475$ haloes in the true field, it slightly surpasses the $404$ well-resolved haloes in the true field. The deviations identified are likely an effect of the specific data choice of halo fields (one mass-weighted and one counts only) and the loss function in Eq~\eqref{eq:mse}. We leave the investigation on more optimally selected mass weighting schemes and the inclusion of more mass bins, both shown beneficial for information extraction in \citet{Hamaus2010} and \citet{Liu2021}, to future studies. In section~\ref{sec:discuss_data_likelihood}, we further discuss the importance of choosing appropriate data vectors.

\subsubsection{Halo velocities}
In the current reconstruction, only the position and mass information of haloes were used in the data. As the full halo catalogue provided by \texttt{AHF} contain additional properties that have not been used as constraints, we can check how well these have been reconstructed. In Fig.~\ref{fig:halo_vel}, we show the velocity dispersion $\sigma_v$ as well as the peculiar velocity $V=(V_x^2+V_y^2+V_z^2)^{1/2}$ for all haloes above the mass resolution limit. While it is evident that the reconstruction contains slightly more haloes, it demonstrates a strong overall alignment with the true distribution in both velocity properties. To further increase the alignment between these distributions, one could consider adding velocity information to the data (also see the discussion in section~\ref{sec:discuss_data_likelihood}). 

\subsubsection{Higher-order statistics}
\label{sec:higher}
As the quality of cosmological data from large-scale galaxy redshift surveys improves, the need for statistical tools capable of extracting more non-linear information is needed. While the power spectrum is still an important probe for Gaussian information contained in the large-scale structure, higher $N$-order statistics are essential to extract non-Gaussian information. We therefore choose to compute the third order statistics, namely the bispectrum, also using \texttt{PYLIANS} \citep{Pylians}. 

The bispectrum captures spatial correlations in the matter distribution over various triangular configurations. In Fig.~\ref{fig:recon_bispec}, we display two different configurations of the reduced bispectrum $Q$, which has a weaker dependence on cosmology and scale, and is defined as
\begin{equation}
    Q\left(k_1, k_2, k_3\right) \equiv \frac{B\left(k_1, k_2, k_3\right)}{P_1 P_2+P_1 P_3+P_2 P_3},
\end{equation}
where $P_i \equiv P(k_i)$ is the power spectrum at $k_i$, $k_i$ are the three wave modes corresponding to the three sides of the triangle, and $B$ is the bispectrum. 

\begin{figure*}
    \centering
    \includegraphics[width=1.0\linewidth]{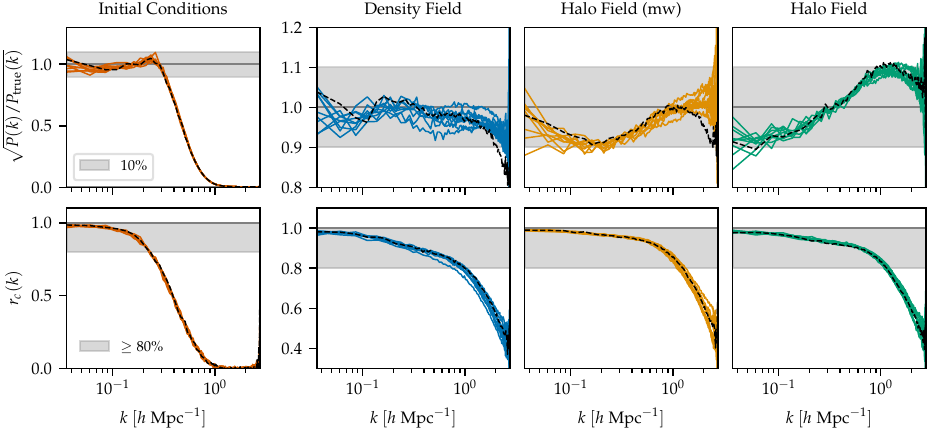}
    \vspace{-0.55em}
    \caption{We quantify the reconstruction quality from ten different optimizations with the transfer function (top) and cross-correlation (bottom). The data in each case is generated from new initial conditions and the corresponding halo catalogue, from which the initial conditions are reconstructed with \texttt{LULO} with $22$ optimization steps in all cases. We display the reconstruction from section~\ref{sec:reconstructions} in dashed black. Apart from minor fluctuations, we demonstrate accurate and consistent recovery of the initial conditions, density field and halo fields for all ground truths.}
    \label{fig:various_data}
\end{figure*}

Two different triangular configurations typically used \citep[see e.g.][]{Jamieson2022b,Doeser2024,Jamieson2024,Bartlett2024} are chosen to asses the reconstructions. First, we express the bispectrum through the magnitude of two wave vectors, chosen to $k_1=0.1h$ Mpc$^{-1}$ and $k_2=1.0h$ Mpc$^{-1}$ as this captures both large and small scale information, and the separation angle $\theta$, i.e. $Q\left(k_1, k_2, \theta \right)$. Second, the bispectrum is expressed through the magnitude of equal wave vectors forming an equilateral triangle, which enables analyzing the fields at all scales. We compute these two forms of the reduced bispectrum for the final over-density field, the mass-weighted halo overdensity and the halo overdensity. Notably, we remove all haloes below the mass resolution limit (see Fig.~\ref{fig:reconstruction_hmf}) before computing the bispectra. In section~\ref{app:bispec_wnf}, we further check the bispectrum obtained for the reconstructed initial conditions.

We not only compare the reconstructed results with the ground truth but also assess the compatibility with the underlying cosmological model by displaying cosmic variance. Using the $1000$ simulations that were used to create the training dataset, we estimate the expected variance in the bispectra. Our reconstructions demonstrate consistency with these predictions as well as the ground truth across most scales. The largest discrepancies are observed at large scales, which is expected since cosmic variance is dominating at those scales. On the other hand, we achieve highly accurate recovery at small scales, particularly for the halo overdensity. In summary, these results demonstrate the capability of \texttt{LULO} to accurately reconstruct non-linear density fields and haloes, despite the challenges posed by a non-differentiable setup.

\subsection{Application on other ground truths }
\label{sec:various_data}

Up to this point, we have only applied \texttt{LULO} on one simulated halo catalogue to reconstruct the underlying cosmic initial conditions. In this section, we test the generalizability of our method by applying it to $10$ other halo catalogues. To generate these, we vary the underlying ground truth initial conditions while keeping the same simulation setup and cosmological parameters. For each new initial conditions, we simulate the corresponding new halo catalogues, which are then used to generate the data in the form of the two halo fields. 

In Fig.~\ref{fig:various_data}, we quantify the reconstructions after $22$ optimization steps for each run (requiring between $93$ and $97$ simulations) in terms of the transfer function and cross-correlation. By also comparing the reconstructions with the results from the previous section, we demonstrate that our method accurately and consistently traverses high-dimensional space to estimate the initial conditions from different halo catalogues. We note that the reconstructions all show similar behaviour, such as slightly undershooting the halo overdensity field $\boldsymbol{\delta}_{\mathrm{count}}^{\mathrm{halo}}$ at $k \sim 0.2h$ Mpc$^{-1}$ and overshooting the mass-weighted halo overdensity $\boldsymbol{\delta}_{\mathrm{mw}}^{\mathrm{halo}}$ at $k \sim 1h$ Mpc$^{-1}$. 

\begin{figure*}
    \centering
    \includegraphics[width=1.0\linewidth]{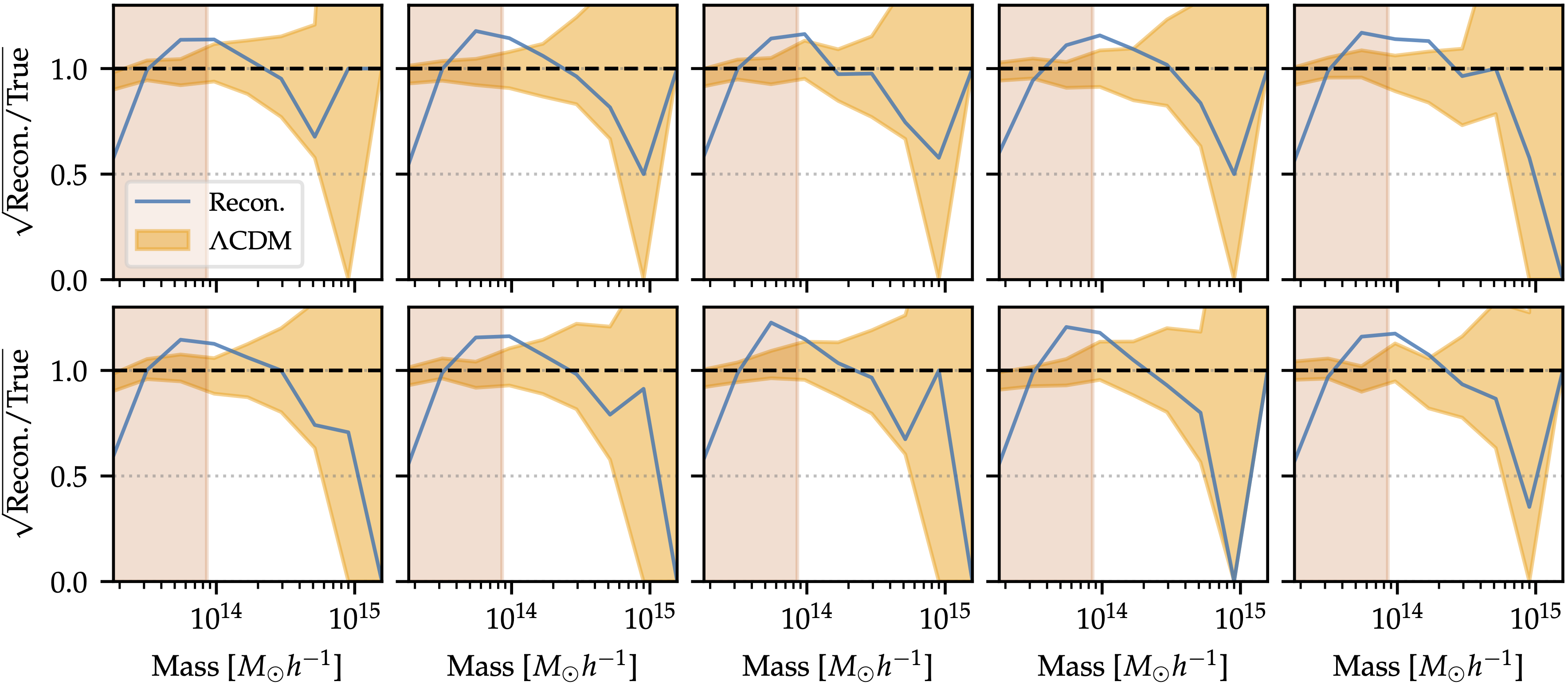}
    \caption{Reconstructions of the halo mass function using different ground truth data. The method tends to predict fewer low- and high-mass haloes and a higher number of intermediate-sized haloes than the respective ground truths. In most optimizations and for most masses, the final halo mass function agrees with the underlying theory prediction of $\Lambda$CDM (yellow band) above the mass resolution limit (red band).}
    \label{fig:various_data_hmf}
\end{figure*}

In Fig.~\ref{fig:various_data_hmf}, we further show the reconstructed halo mass functions as compared with the respective truth and the $\Lambda$CDM scatter as estimated from the $1000$ training simulations (see section~\ref{sec:halo_reconstruction}). \texttt{LULO} consistently recovers halo mass functions that agree with the theory, while typically under-predicting the number of low-mass haloes ($M<0.4\times 10^{14}M_{\odot}h^{-1}$; below mass resolution limit) and high-mass haloes ($M>3\times10^{14}M_{\odot}h^{-1}$) and over-predicting the number of low-intermediate mass haloes. To improve the reconstruction quality, future work will focus on refining the architecture, training strategy, and data likelihoods (see discussion in section~\ref{sec:discuss_data_likelihood}). 

\section{Discussion}
\label{sec:disc}
The last decade has seen significant advancements in modelling the observed galaxy distribution. However, current methods often face challenges when incorporating non-linear scales due to the complexity of the required physics simulators to model the data. Our novel learning-to-optimize approach allows for the direct inclusion of complex and accurate non-linear physics models without requiring differentiability. In the following, we extend the discussion on the importance of non-linear modelling and how to optimally select loss or likelihood functions to use most of the data. 

\subsection{The importance of modelling non-linear scales}
\label{sec:discuss_non_linear}

Non-linear modelling is essential because the observed objects are inherently non-linear, having formed through gravitational collapse. As we have also demonstrated, incorporating non-linear scales is crucial for accurately recovering the large-scale linear initial conditions, aligning with the findings in \citet{Doeser2024}. In the context of inferring the initial conditions from galaxy clustering data, as discussed in \citet{Stopyra2023}, a highly accurate non-linear model is required during inference to obtain accurate posteriors of present-day structures. 


Reconstructions of the initial conditions of our Universe can be leveraged to run constrained simulations that evolve primordial fluctuations to the particular observable large-scale structure distribution of galaxies \citep[e.g.][]{Lavaux_2010,Leclercq2015,Wang2016,Sorce2018,Hutt2022,Mcalpine2022,mcalpine2025manticoreprojectidigital}.

These simulations enable investigating how structures in our observable Universe formed, thus creating a laboratory to search for model consistencies between $\Lambda$CDM and observations. Recent work includes, for example, analysing the spatial distribution of elliptical galaxies and spiral galaxies \citep{Sawala2023}, intricate properties of clusters \citep{Sorce2018,Jasche2019,Hutt2022,Mcalpine2022,mcalpine2025manticoreprojectidigital}, variance of halo properties \citep{Stiskalek2024}, modified gravity \citep{Bartlett2021}, galaxy intrinsic alignment \citep{Tsaprazi2021}, large-scale environment of supernovae \citep{Tsaprazi2022}, gamma-ray emission from dark matter particle annihilation \citep{Kostic2023}, magnetic fields on cosmological scales \citep{Hutschenreuter2018}, and massive neutrinos \citep{Elbers2023}. There have also been dedicated efforts targeting the Local Group, our immediate cosmic neighborhood, to model its detailed structure and dynamics \citep[see e.g.][]{Gottl,Libeskind2020,Sawala_2021,Wempe2024}, as well as studies focusing on the formation, evolution, and fate of galaxies and galaxy clusters located further away from us, at higher redshifts \citep{Ata2022,Byrohl2024}. One may further note that re-simulations of our observable Universe provide access to the complete dynamical evolution of non-linear peculiar velocities, which are essential for peculiar velocity corrections in both Hubble constant measurements using supernova data \citep[e.g.,][]{Kenworthy2022, Riess2022, Peterson2022} or using gravitational wave sources \citep{Mukherjee2021}. 

As an increasing number of scientific objectives rely on cosmic initial conditions, improving their accuracy through enhanced non-linear data modelling is becoming critical. In this work, we advance non-linear modelling by incorporating non-differentiable components in the form of a halo-identification algorithm, enabling more accurate and reliable reconstructions of initial conditions.

\subsection{Choosing appropriate data likelihoods}
\label{sec:discuss_data_likelihood}

Throughout this work, our focus has been on optimizing and refining the search mechanism for accurate reconstructions using non-linear and non-differentiable physics simulators. Entering the non-linear regime also necessitates careful modeling of the loss functions and likelihoods. Specifically, in this study, where halo catalogs are used as data, we have explored approaches to model count data. Previous studies have examined the additional information provided by mass-weighted halo count fields as compared to count fields \citep{Seljak2009,Hamaus2010,Liu2021}. In this work, we chose to model the halo count data using both a halo overdensity (with counts only) and a mass-weighted halo overdensity, and applying a Gaussian likelihood between the reconstruction and the data. 

\begin{figure*}
    \centering
    \includegraphics[width=1.0\linewidth]{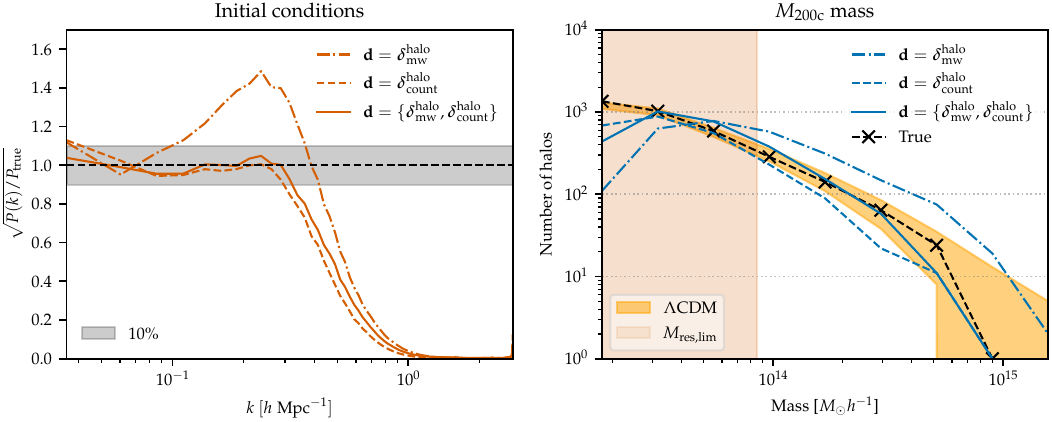}
    \vspace{-0.65em}
    \caption{Reconstructions using different halo fields as data $\mathbf{d}$, as generated by the same halo catalog. By only using either the mass-weighted halo overdensity $\boldsymbol{\delta}_{\mathrm{mw}}^{\mathrm{halo}}$ or halo overdensity $\boldsymbol{\delta}_{\mathrm{count}}^{\mathrm{halo}}$ in the data likelihood, the initial conditions (left), and hence the present-day halo mass function (right), are not accurately reconstructed. In particular, only using the mw-field results in an overestimation of high-mass haloes and hence a larger power in the initial conditions. On the contrary, using the count overdensity results in accurately recovering only the most numerous (low-mass) haloes, and a slightly lower power in the initial conditions. Using both the halo counts as well as their masses enables a more accurate reconstruction of the initial conditions and, in turn, the halo mass function.}
    \label{fig:mw_vs_nw}
\end{figure*}

To evaluate the individual contributions of the halo overdensity and mass-weighted halo overdensity, we conducted experiments using each field independently, both for training the neural optimizer and as input data during optimization. The reconstructed halo mass functions for these cases are presented in Fig~\ref{fig:mw_vs_nw}. We observed that omitting either the count or mass constraints leads to worse reconstructions. For instance, in the case of only using the mass-weighted halo overdensity, we see that the power of the reconstructed initial conditions overshoots the true cosmological power, thus prioritizing the creation of massive objects. While using only the mass-weighted field resulted in reconstructing more high-mass haloes, using only the halo overdensity resulted in reconstructing more of the numerous low-mass haloes. 

Incorporating both counts and mass breaks these degeneracies, enabling more accurate reconstructions. Similarly, a recent study by \citet{Bayer2023b} demonstrated that adding halo velocity data alongside halo positions further enhances the reconstruction of initial conditions. These findings highlight the potential of incorporating additional information to achieve even more accurate reconstructions. 

While halo or galaxy clustering serves as one key data source, from which we have demonstrated the importance of careful modelling, other cosmological data sources can also be considered. Field-level inference or reconstruction of initial conditions has been explored using various datasets, including Lyman-alpha forests \citep{Horowitz2019,Horowitz2022,Porqueres2020,Porqueres2021}, cosmic shear measurements \citep{Porqueres2021,Porqueres2022,Porqueres2023}, photometric redshifts \citep{Tsaprazi2023}, and peculiar velocity tracers \citep{Prideaux-Ghee2022}. Our approach naturally allows for the simultaneous incorporation of multiple sources of cosmological data, which offers a pathway to more comprehensive and accurate reconstructions of initial conditions in the future.

\subsection{Flexibility and computational performance of \texttt{LULO}}
\label{sec:compcost}

We emphasize that our framework \texttt{LULO} is compatible with any simulation pipeline. Alternative forward models offering other levels of fidelity or involving different output quantities to match new datasets can be seamlessly integrated as long as input-output pairs can be generated. The computational cost therefore depends on the desired accuracy and can be adjusted to prioritize speed with approximate simulators or higher fidelity at an increased computational expense.

In this work, to demonstrate the successful integration of non-differentiable models without requiring modifications to the simulators, we used the \texttt{Gadget-IV} code \citep{Springel2021} and the \texttt{AHF} halo identification algorithm \citep{Gill2004, Knollmann2009}. The combined computational cost of these with $128^3$ simulated particles within a $250h^{-1}$ Mpc box was approximately $5.5$ CPU-hours per simulation, leading to a total of around $27.5$k CPU-hours for generating the training data. The exact number of simulations required for the training data was not explored in this work but remains an area of interest for future investigations. For this setup and having $5000$ training pairs, a single model required approximately $120$ GPU-hours for training.

During optimization, as also visualized in Figure~\ref{fig:opt_process}, each step requires on average $4.3$ simulations for the line search. In total, calling the simulation pipeline approximately $96$ times is needed for $22$ optimization steps, which corresponds to a total wall time of $11$ hours. This may be a conservative estimate, as convergence is approached after approximately $30$ simulations as seen in Fig.~\ref{fig:opt_process}. We note that predicting the update direction on the GPU takes less than $2$ seconds per optimization step, making this cost negligible in comparison to the simulation costs.

In future work, higher-resolution simulations could be incorporated to capture finer cosmological structures. Forward models that include hydrodynamical or other dynamical processes could also be integrated. Field-level emulators, such as the ones developed by \citet{Jamieson2022b,Jamieson2024} for simulating non-linear structure formation, could also be leveraged. As demonstrated in \citet{Doeser2024} within Bayesian inference of initial conditions, these emulators can reduce computational costs by more than a factor of $100$ compared to full $N$-body simulations, with only a minor trade-off in accuracy. The flexibility of \texttt{LULO} to incorporate the latest advancements from the community in structure formation and galaxy biasing modelling, without requiring differentiability, ensures it remains adaptable to evolving methodologies.

\section{Conclusions}
\label{sec:conc}

With next-generation galaxy surveys approaching, accurate data models are crucial for maximizing information from small-scale clustering. Extracting cosmological insights at non-linear scales is particularly challenging due to the complex non-linear modelling of large-scale structures and the relationship between the observed galaxies and the underlying dark matter distribution, often involving numerical simulators and non-differentiable data models. To tackle these challenges, we develop \textit{Learning the Universe by Learning to Optimize} (\texttt{LULO}), a framework for reconstructing 3D cosmic initial conditions by fitting state-of-the-art non-differentiable simulators to cosmological data at the field-level. 

By fitting explicit physics models that span from the initial conditions to the data, field-level inference surpasses traditional approaches based on statistical summaries while also enabling detailed causal analysis \citep[see e.g.][]{Jasche2019,Mcalpine2022,Wempe2024,mcalpine2025manticoreprojectidigital}. While most field-level approaches rely on differentiable physics simulators, we follow a complementary approach that goes beyond the requirement of differentiability by accelerating the optimization algorithm itself. \texttt{LULO} also separates the neural optimization pipeline from the physics simulator, allowing any simulator to be integrated without extra development.

As opposed to recent uses of deep learning that directly predicts the input or output of a physics model, our neural optimizer is tasked with learning the dynamical behaviour of a simulator by learning how differences between the data and the model prediction map to updates in the initial conditions. During optimization, the optimizer is then tasked to predict a search direction in which the initial conditions should be updated to minimize the data discrepancy. Specifically, our framework keeps high-fidelity physics simulators in the loop, allowing for ongoing validation of the updated initial conditions throughout the process and providing full explainability, thus addressing common remarks of black-box deep learning \citep[e.g.][]{Angulo2021, Huertas-Company2022}. 

In this work, we have demonstrated that \texttt{LULO} accurately reconstructs the initial conditions from $M_{200\mathrm{c}}$ halo catalogues. These catalogues are generated using the high-fidelity, dark-matter-only simulation code \texttt{Gadget-IV} and the spherical overdensity algorithm \texttt{AHF}. In particular, we show that the transfer function and cross-correlation of the initial conditions are accurately recovered in the linear regime. When forward simulated, we recover non-linear structures at the present epoch to scales $k\gtrsim1h$ Mpc$^{-1}$. Our work highlights the critical role of mode coupling in gravitational structure formation for modelling non-linear structures to $k\sim1h$ Mpc$^{-1}$ at the present epoch. Accurately describing linear scales ($k < 0.1h$ Mpc$^{-1}$) in the initial conditions, where it can be compared to linear theory, thus necessitates accounting for non-linear scales.

In summary, our approach integrates machine learning as an optimizer while preserving full physics simulations in the loop. By streamlining the development cycle and enabling the use of any high-fidelity physics simulator, our approach ensures both high scalability and physical fidelity. While our results demonstrate the potential of \texttt{LULO} for field-level inference, several challenges remain. Future work will explore its scalability to larger simulations, generalization to different physics models, and inclusion of observational systematics and survey characteristics in the modelling pipeline. In conclusion, \texttt{LULO} demonstrates a powerful framework for leveraging complex, non-differentiable simulators to model next-generation cosmological data at non-linear scales, paving the way for more precise and comprehensive cosmological field-level inference.

\section*{Acknowledgements}
We thank Stuart McAlpine, Matt Ho, and Guilhem Lavaux for useful discussions related to this work and their feedback on the manuscript. We also thank Richard Stiskalek and Stephen Stopyra for useful discussions related to the setup of the IC generator and the halo finder algorithm, respectively. This research utilized the Sunrise HPC facility supported by the Technical Division at the Department of Physics, Stockholm University. The computations were also enabled by the Berzelius resource provided by the Knut and Alice Wallenberg Foundation at the National Supercomputer Centre at Linköping University, Sweden. This work has been enabled by support from the research project grant ‘Understanding the Dynamic Universe’ funded by the Knut and Alice Wallenberg Foundation under Dnr KAW 2018.0067. JJ acknowledges the hospitality of the Aspen Center for Physics, which is supported by National Science Foundation grant PHY-1607611. The participation of JJ at the Aspen Center for Physics was supported by the Simons Foundation. JJ acknowledges support by the Swedish Research Council (VR) under the project 2020-05143 -- ``Deciphering the Dynamics of Cosmic Structure". This work was supported by the Simons Collaboration on “Learning the Universe”.

\section*{Data Availability}
The data underlying this article will be shared on reasonable request to the corresponding author.


\bibliographystyle{mnras}
\bibliography{ref}

\appendix

\section{Optimization Guarantees}
\label{app:opt_guarantees}
For convex and certain quasi-convex high-dimensional problems, \citet{Golovin2020} showed that a random walk optimization strategy will converge within an $\epsilon$-ball around the optimum in a finite number of function evaluations. We draw inspiration from their geometrical perspective to claim that walking in the correct half-space is sufficient to reach the optimum. \\

\noindent
\textbf{Theorem}: If $f$ is a convex function and the search direction $d$ points to the correct half-space, then a more optimal solution can be found by performing a line search along $d$.

\noindent
\begin{proof}
Let $x$ be the current iterate, $d$ be the search direction, and $\gamma > 0$ be the step size. Consider the function $g(\gamma) = f(x + \gamma d)$, which is a convex function of $\gamma$ since $f$ is convex. By convexity, we have
\begin{equation}
g(\gamma) \leq g(0) + \gamma g'(0),
\label{eq:app_1}
\end{equation}
where $g'(0)$ is the derivative of $g$ at $\gamma=0$. Note that $g(0) = f(x)$ and $g'(0) = \nabla f(x)^\top d$, since $g$ is just the restriction of $f$ to the line $x + \gamma d$. Substit
uting these values into Eq.~\eqref{eq:app_1}, we get
\begin{equation}
f(x + \gamma d) \leq f(x) + \gamma \nabla f(x)^\top d.
\label{eq:app_2}
\end{equation}
If $d$ points to the correct half-space, then $\nabla f(x)^\top d > 0$. Thus, for $\gamma>0$, the second term on the right-hand side of Eq.~\eqref{eq:app_2} is positive, which means that $x+\gamma d$ is a better solution than $x$. Therefore, we can perform a line search along $d$ to find the optimal solution.
\end{proof}

\section{Approximate exact line search}
\label{app:line_search}
As each function evaluation in the line search can be computationally heavy – as in our case with a full simulation – we aim for a minimal amount of evaluations. We pick an effective line search algorithm called Approximate Exact Line Search (\texttt{AELS}), which is an adaptive variant of golden section search that comes with convergence guarantees \citep{Fridovich-Keil2020}. Applying \texttt{AELS}, as seen in Appendix~\ref{app:towards_opt} and Fig.~\ref{fig:opt_process} in particular, requires on average $4.3$ simulations per line search.

\renewcommand{\thefigure}{ C\arabic{figure}}
\begin{figure*}
    \centering
    \includegraphics{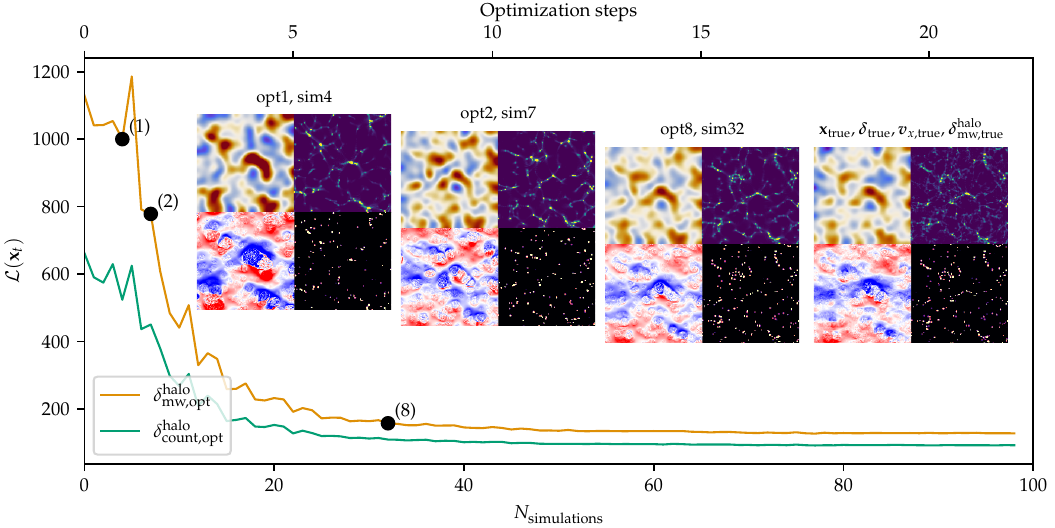}
    \vspace{-1em}
    \caption{The data discrepancy, as measured by the mean squared error $\mathcal{L}(\ve{x}_t)$ defined in Eq~\eqref{eq:mse} between the data and the forward simulated initial conditions in terms of the mass-weighted halo overdensity $\boldsymbol{\delta}_{\mathrm{mw,opt}}^{\mathrm{halo}}$ and the halo overdensity $\boldsymbol{\delta}_{\mathrm{count,opt}}^{\mathrm{halo}}$, is decreasing as a function of optimization steps. At each optimization step (upper x-axis), the line search requires a certain number of simulations (lower x-axis). The non-monotonic decrease in the data discrepancy results from the line search, which requires a few simulations to find the step size in the given search direction. In monitoring the improved reconstructions, we show for a few steps the initial conditions (upper left), and the corresponding forward simulation output at $z=0$: dark matter overdensity (upper right), Lagrangian velocity field (lower left), and mass-weighted halo field (lower right). Note the optimization is the same as in section~\ref{sec:reconstructions} but here with new 2d-slices. A visual comparison with the data $\mathbf{d}$ (right-most inset) shows similarities already after one and two optimization steps. Although convergence is approached after $\sim 8$ steps, we let it fully converge and run $22$ optimization steps ($96$ simulations).}
    \label{fig:opt_process}
\end{figure*}

\renewcommand{\thefigure}{ C\arabic{figure}}
\begin{figure}
    \centering
    \includegraphics[width=1.0\linewidth]{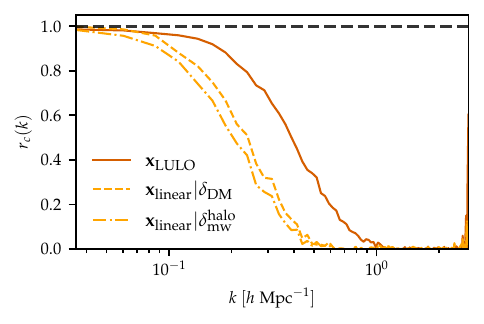}
    \caption{Cross-correlation between the ground truth and the reconstructed initial conditions from our non-linear reconstruction algorithm \texttt{LULO} compared to the optimal linear reconstruction. Using the two halo fields as input, \texttt{LULO} shows a higher cross-correlation than the linear method, here using either the dark-matter field or the mass-weighted halo overdensity as data.}
    \label{fig:WF_recons}
\end{figure}

\renewcommand{\thefigure}{ C\arabic{figure}}
\begin{figure*}
    \centering
    \includegraphics[width=1.0\linewidth]{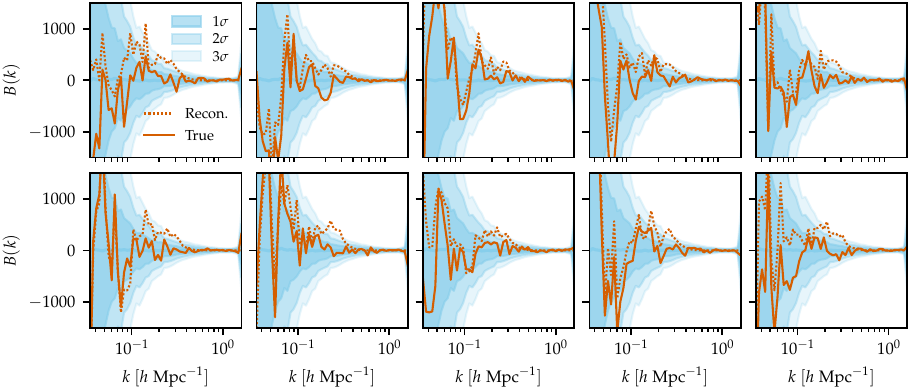}
    \caption{Equilateral bispectrum of the reconstructed initial conditions compared to $1000$ random Gaussian fields and the ground truth. The $1$-, $2$-, and $3\sigma$ deviations are computed from these random fields and shown in shaded blue. In all $10$ cases (see Section~\ref{sec:various_data}), the reconstructions show good overall agreement with Gaussian expectations and largely match the respective true initial conditions.}
    \label{fig:bispec_wnf}
\end{figure*}

As part of the development, we also used the bisection bracketing method to benchmark and verify the slightly higher performance of the AELS method. We highlight that AELS works without specifying a search interval and, while being approximate, can be tuned to satisfy accuracy requirements. The hyperparameters that need to be chosen are the initial step size guess $T$, and the adjustment factor $\beta$. We note that having the line search method applied at each optimization step requires setting $T$ at each iteration $t$. We find empirically that $T_{t+1} = \gamma_t$, where $\gamma_t$ is the step size chosen by the algorithm at iteration $t$, if $\alpha < 1$ (see Algorithm 2 in \citet{Fridovich-Keil2020}) and $T_{t+1} =(\beta^2 \gamma_t + T_t)/2$ if $\alpha > 1$ are effective choices. We initialize with $T_0=0.01$ and $\beta = 0.7$, which slightly differs from the $\beta = 2/(1+\sqrt{5})$ chosen in \citet{Fridovich-Keil2020}. 

Increasing the value of $\beta$ results in a lower exploration efficiency but a heightened level of sensitivity. We found that a higher $\beta$ together with the addition of a patience parameter $p_{\mathrm{LS}}$, which stops the line search if a $f_{\mathrm{LS}}$ fractional improvement of the data discrepancy has not occurred for $p_{\mathrm{LS}}$ iterations, worked better for our purposes. Similarly, between the optimization steps, we introduce $p_{\mathrm{OPT}}$ and $f_{\mathrm{OPT}}$ that checks for fractional improvements. We picked $p_{\mathrm{LS}}=3$, $p_{\mathrm{OPT}}=5$, and $f_{\mathrm{LS}}=f_{\mathrm{OPT}}=0.001$.

\section{Reconstruction performance}

In this section, we further examine our reconstruction algorithm \texttt{LULO}. In section~\ref{app:towards_opt}, we present the intermediate reconstruction results at each optimization step. In section~\ref{app:init}, we discuss different initialization strategies. To show the high quality of the final reconstruction, we compare our result with the optimal linear reconstruction in section~\ref{app:linear_recon}. Finally, in section~\ref{app:bispec_wnf}, we show the bispectrum of the reconstructed initial conditions.

\subsection{Towards optimum}
\label{app:towards_opt}

In Fig.~\ref{fig:opt_process}, we monitor the mean squared error of the mass-weighted and the count-weighted halo fields with their respective ground truths as a function of the number of simulations (lower x-axis) and the number of optimization steps (upper x-axis). The figure displays the optimization process leading to the reconstruction analyzed in seciton~\ref{sec:reconstructions}. On average, each line search within one optimization step requires $5$ simulations. Only $4$ optimization steps, corresponding to $20$ simulations, are required to reduce the error, after which further optimization only marginally improves the reconstructions. 

In the inset figures in Fig.~\ref{fig:opt_process}, we display four different fields and their improvement as a function of optimization steps. In the upper left, we show the initial conditions smoothed over $5$ voxels ($\sim10h^{-1}$ Mpc) with a Gaussian kernel. Because of temporal mode coupling between large scales in the initial conditions and small scales in the data, to be discussed in more detail in section~\ref{sec:reconstructions}, a high correlation between the reconstruction and the truth down to roughly $10h^{-1}$ Mpc is sufficient to obtain forward predictions that match the data at non-linear scales of $\sim 1h^{-1}$ Mpc. This is seen in the lower right plot, where we show the mass-weighted halo field. Notably, the sole objective of the optimizer is to update the initial conditions such that the data discrepancy for the halo fields is reduced. As the number of optimization steps taken increases, the visual alignment increases.

\subsection{Initialization and generalizability}
\label{app:init}
While we leave tests of other initialization strategies to future studies, one may note that initializing from a zero field results in the first mapping being $\Delta \ve{x} = \ve{x}_{\mathrm{true}}$ and $\Delta \ve{d} = \ve{d}$. This setup effectively tasks the neural optimizer with estimating the true initial conditions directly from the data. This resembles e.g. \citet{Shallue2023,Jindal2023,Chen2023,Legin2023} who use neural networks to map non-linear data to the initial density field. These predictions could be used as the initial guess in our framework. We stress, however, the fundamental difference between these neural models and our neural optimizer, which does not aim to make any predictions directly but to propose a search step in the initial conditions space given the data discrepancy. This enables always testing the proposed initial conditions within the full physics model and computing the new data likelihood, before providing the next update direction. 

We note that zero initialization falls outside the range of the training dataset. The neural optimizer has only been trained on data differences originating from pairs of initial conditions each having unit variance. While the power mismatch is addressed through the line search, the neural optimizer determines the update direction, a task it performs with high accuracy even in these cases. It thus demonstrates a general understanding of the underlying dynamical behaviour, which results in efficient searches through high-dimensional spaces. 

\subsection{Comparing to linear reconstruction}
\label{app:linear_recon}

To demonstrate the importance of modeling non-linear scales, we compare our reconstruction quality as measured by the cross-correlation with traditional linear methods. Following \citet{Legin2023}, we compare our non-linear reconstruction with the optimal linear reconstruction obtained through Wiener filtering. This comparison, shown in Fig.~\ref{fig:WF_recons}, illustrates the improvement achieved by modelling non-linear scales in the data to reconstruct the initial conditions.

A linear reconstruction algorithm assumes that the observed data vector $\ve{d}$ is related to the true signal $\ve{x}$ (in our case, the initial conditions) via a linear operator $\mathbf{R}$ and additive noise $\ve{\varepsilon}$ as
\begin{equation}
\ve{d} = \mathbf{R}\ve{x} + \ve{\varepsilon}.
\end{equation}
In our context, $\mathbf{R}$ corresponds to convolving the initial white-noise field $\ve{x}$ with the primordial power spectrum, evolving it to redshift $z = 0$ using the transfer function, and applying a linear bias to model the halos. Under this linear model and assuming Gaussian statistics, the optimal estimator \citep[see e.g.][]{Zaroubi_1995} is the Wiener filter:
\begin{equation}
\mathbf{F} = \langle\ve{x}\ve{d}^{\dagger}\rangle\langle\ve{d}\ve{d}^{\dagger}\rangle^{-1} = \frac{\mathbf{S}\mathbf{R}}{\mathbf{N} + \mathbf{S} \mathbf{R}^\top \mathbf{R}},
\end{equation}
where $\mathbf{S}$ and $\mathbf{N}$ are the signal and noise covariance matrices. The linear reconstruction is then given by
\begin{equation}
\hat{\ve{x}}_{\mathrm{linear}} = \mathbf{F} \ve{d},
\end{equation}
with each element of $\mathbf{F}$ satisfying $\mathbf{F}_{ij} \geq 0$, since covariance matrices are positive semi-definite and the filter cannot amplify signal power.

The cross-correlation between the true initial conditions and the linear reconstruction is thus fundamentally limited by the cross-correlation between the data and the true signal as
\begin{equation}
\langle\ve{x}_{\mathrm{true}} \hat{\ve{x}}_{\mathrm{linear}}^{\dagger}\rangle = \mathbf{F} \langle\ve{x}_{\mathrm{true}} \ve{d}^{\dagger} \rangle \leq \langle\ve{x}_{\mathrm{true}} \ve{d}^{\dagger} \rangle.
\end{equation}
We display this optimal cross-correlation achieved through linear reconstruction in Fig.~\ref{fig:WF_recons}. Our non-linear reconstruction with \texttt{LULO} yields a higher cross-correlation, highlighting the importance of non-linear modelling.

\subsection{Bispectrum of reconstructed initial conditions}
\label{app:bispec_wnf}
Our reconstruction, performed without an explicit prior, yields a maximum likelihood estimate of the true initial conditions and faithfully captures the large-scale structure and statistical properties where the data is informative. As shown in the power spectrum in Fig.~\ref{fig:reconstruciton_powspec} and Fig.~\ref{fig:various_data}, the reconstructed initial conditions exhibit Gaussian random behaviour, characterized by a flat spectrum, down to a certain scale.

Similarly, the reconstructed bispectrum is expected to be consistent with that of a Gaussian white-noise field for certain scales. To assess this, we compare the bispectrum of the reconstructed initial conditions to that of random Gaussian fields. In particular, we compute the equilateral bispectrum for $1000$ random Gaussian fields and show the comparison between these, the ground truth, and the reconstructions in Fig.~\ref{fig:bispec_wnf} for the $10$ different ground truths (see section~\ref{sec:various_data}). Note, we do not display the reduced bispectrum here as the power spectrum of the reconstructed initial conditions goes to zero for smaller scales (see Fig.~\ref{fig:reconstruciton_powspec}), and the reduced bispectrum would therefore not be informative. We find good overall agreement, with the reconstructed bispectrum closely matching Gaussian expectations, while noting a consistent deviation around $k=0.2$-$0.3h$ Mpc$^{-1}$. At both larger and smaller scales the bispectrum aligns well with Gaussian predictions and largely agrees with that of the true initial conditions.

\end{document}